\documentclass{ieeetran}

\usepackage{graphics}
\begin{document}

% Added on 15au18:

\onecolumn

% \emph{P O Box 32104 Panama City, FL 32407-8104} matadorsalsa at runbox dot com, 964 S. Shannon Ave. Indialantic, FL 32903. This is personal research. Otherwise, email the author at michael.c.kobold at navy dot mil.

% \title{Observations of nonlinear eigen-states}

%\title{Large spot-size observations \\ of nonlinear eigen-states
%\mbox{\LARGE{modal analysis of clattering hardware, theory and data}}}

\title{Remote vibrometry recognition \\ of nonlinear eigen-states
\mbox{\LARGE{for object coverage of randomly large size}}}

% \subtitle{modal analysis of clattering hardware via sensing with large spot size, theory and data}

% For IEEE need to remove \subtitle, \institute, ...

%\subtitle{Including modal response of large spot size return from clattering contact structures}
%%\author{Michael C. Kobold\inst{1} \and Michael McKinley % etc
%%\author{Michael C. Kobold\inst{1} \and Second author\inst{2}% etc

% \author{Michael C. Kobold\inst{1} % etc  % But we are not using institution

%\author{Michael C. Kobold % etc
%\thanks{Dr. Pedro Encarnac\'ion of Colorado Springs and especially Michael McKinley of Arlington, Texas provided much appreciated assistance, www .linkedin .com /in/michaelmckinley2001 (remove spaces to activate link).}%
%}                     % Do not remove

\author{\IEEEauthorblockN{Michael C. Kobold}
(\IEEEauthorblockA{Indialantic, Florida})  \\
and
\IEEEauthorblockN{Michael C. McKinley}
(\IEEEauthorblockA{Arlington, Texas})
\thanks{\emph{M. Kobold, P.E., 964 S. Shannon Ave. Indialantic, FL 32903}, worksite location: P O Box 32104 Panama City, FL 32407-8104.}
\thanks{matadorsalsa @ runbox . com, michael.c.kobold @ navy . mil} \\
\thanks{\emph{M. McKinley, independent consultant, www .linkedin .com/in/michaelmckinley2001 }(Remove spaces to activate links and email addresses.)}
\thanks{The helpful comments of Adam Al-Saleh of Panama City, Florida are gratefully appreciated.}
\thanks{The insight and assistance of Dr. Pedro Encarnaci\'on of Colorado Springs are appreciatively and always welcome.}
}

%%\thanks{\emph{Present address:} Insert the address here if needed}%
%
%%\offprints{}          % Insert a name or remove this line
%

% \institute{\emph{P O Box 32104 Panama City, FL 32407-8104} matadorsalsa at runbox dot com, 964 S. Shannon Ave. Indialantic, FL 32903. This is personal research. Otherwise, email the author at michael.c.kobold at navy dot mil.}

%\institute{Insert the first address here \and the second here}
%
\date{Received: 0jan19 / Revised version: 0feb19}

%  Moved before abstract for IEEE

% Add these before the �\maketitle� command (you may also take a look here for more details. Also another hint : remove the \thispagestyle{plain} command which is right after maketitle as it suppresses the copyright line).

\IEEEoverridecommandlockouts
\IEEEpubid{\makebox[\columnwidth]{978-1-4799-7492-4/15/\$31.00~
\copyright2015
IEEE \hfill} \hspace{\columnsep}\makebox[\columnwidth]{ }}

\maketitle

%
%\abstract{

%\begin{abstract}
%The effect of an observer on measurement of small scale energy and transition events can be substantial, which drives the sensor design.  For objects of greater size such as waves on the sea

\begin{abstract}
For objects of ``large'' vibration size such as waves on the sea surface, the choice of measurement method can create different understandings of system behavior.  In one case, laser vibrometry measurements of a vibrating bar in a controlled laboratory setting, variation in probe spot size can omit or uncover crucial structural vibration mode coupling data.  In another case, a finite element simulation of laser vibrometry measures a nonlinearly clattering armor plate system of a ground vehicle.  The simulation shows that sensing the system dynamics simultaneously over the entire structure reveals more vibration data than point measurements using a small diameter laser beam spot, regardless of the variation of footprint (coverage) boundaries.  Furthermore, a simulation method described herein allows calculation of transition probabilities between modes (change-of-state).  Wideband results of both cases demonstrate the 1/$f$ trend explained within -- that the energy of discrete structural vibration modes tends to decrease with increasing mode number (and frequency), and why.  These results quantify the use of less expensive non-imaging classification systems for vehicle identification using the remote sensing of surface vibrations \emph{while mitigating spectral response distortion due to coverage variation on the order of the structural wavelength} (spectral reduction or elimination).
\end{abstract}

%Since vehicles have been remotely classified with similar modal measurement methods, these results demonstrate confidence for vehicle identification using the remote identification of surface vibrations.
%\end{abstract}

% } %end of abstract


%

% \maketitle
%
\section{Introduction}
\label{intro}

Remote vibrometry has many forms.  The use of optics to sense vibration is a mature technology dates back to methods of observations with knife-edge imaging invented by  Jean Bernard L\'eon Foucault in 1858 that can be used to measure mirror shape.  The method was also used by T\"opler \cite{ppr:toepler1866ueber} in 1866 to measure phase and later used to remove unwanted optical phase effects:  ``\dots the Schlieren method, where all the spectra on one side of the central order are excluded." \cite{book:BornAndWolf} The \emph{Schlieren} method, a German word for \emph{streak}, uses Foucault's knife-edge in the ``Fourier plane" at a distance of a focal length from a collecting lens.  A related method, the streak tube, can optically process an entire slice of underwater scenery nearly instantaneously \cite{ppr:overviewJaffeStrand01, ppr:imagingModelStrand91, ppr:uwEOmineIDSPIEStrand95}.  More recent physics discoveries use vibration to measure atomic size effects. Microscopy resolution smaller than the optical diffraction limit can be accomplished with the Nobel prize winning \cite{misc:scanningTunnelingNobelPrize}) scanning tunneling microscope.  Other microscopes image S-shells of large atoms and, in some atoms, their P-shells.  The microscopic range, remote vibrometry variant, the atomic force microscope (AFM) is able to detect separate molecules by measuring the difference in van der Waals force as a tiny vibrating cantilever sensitive to the atomic outer shell locations, scans over the sample.  These are some examples of remote vibrometry.  This article describes how the nature of observations can provide an different pictures of the vibrational modes of common vehicles and other integrated structures built from plates, shells, and membranes (the ``body-in-white").

\mbox{ }

\emph{SUPPORTING RESULTS:}

Three types of analysis provide support for this investigation of observation methods for structural surface waves.  (i) Experimental results of a driven clamped bar with full fixity at both ends provides a laser vibrometry spectrum for both small and large probe beam spot sizes.  (ii) The author's master's thesis \cite{MSthesis:kobold06} contains calculation results of imaging a vehicle hull at 4 kilometers. Identification of the target vehicles based on spectral fingerprints formed from laser vibrometry.  The target armor clattering against the vehicle hull provides a surface vibration that modulates the probe beam. (iii) Several analytical calculations using explicitly nonlinear mechanics validate the distinct clattering spectrum.  These experimental, simulation, and analytical results provide cross-supporting reinforcement for the three observational characteristics described in this work.

%characteristics described in this paper.  The results of this work are intended to be extended to ocean surface wave analysis \cite{book:TalleyPickard11}; they form a foundation for surface wave analysis in all media.

\mbox{ }

\emph{OBSERVATIONS:}

This work defines three characteristics of composite structures such as vehicles.  (1) The lower frequency modes are nearly discrete.\footnote{In practice, modal engineers see effectively discrete modes for the fundamental and other low frequency modes.  Here \emph{fundamental} means the lowest frequency mode, rather than the music definition as the largest common integer-based factor.  For 300 Hz beat with 500 Hz our modal fundamental mode would be at 300 Hz, not 100 Hz.  The experience with effectively discrete modes occurs within trade secret production engineering analysis tasks such as automotive ride and handling -- one of the most guarded secrets.  These frequencies and mode shapes are rarely published in the literature.} (2) Energy from driven modes flows into other modes. And finally, (3) due to theory and observations defined herein, vibration strain-energy is higher in the fundamental mode, with energy decreasing as modal frequencies increase.  The latter two characteristics can be considered \cite{ppr:Onsager31} to be consequences \cite{book:Gibbs1906} of the second law of thermodynamics \cite{misc:meaningOfLife}.

%%Some vibration modes may not be obvious from return due to a probe beam with a small spot size.  The nature of clattering plates becomes more clear when viewing the entire optical image of the return and processing its spectrum at multiple locations.  Even the non-imaging (spatially integrated radiant flux) spectrum of the probe return from
%%...

The probe beam spot size affects vibration modes that are observed with optics.  Some vibration modes may not be obvious from return due to a probe beam with a small spot size, as discussed with Flight Lieutenant Ngoya Pepela's thesis \cite{MSthesis:NgoyaPepelaLaserVib03} in Figures \ref{allModes} and \ref{missedMode} for a vibrating bar.  In the simulation of clattering armor plates, more mode information is acquired when viewing the entire optical image and processing its spectrum at multiple locations.  In spite of some interference, the non-imaging\footnote{Imaging systems require more hardware to calculate pixel values.  Non-imaging systems integrate power incident on the entire optical aperture.  This power is termed the radiant flux, measured in watts.  It is the flux of the Poyting vector $\vec{S}$ through an enclosing surface $ \Phi_{\mbox{\scriptsize{E}}} = \oint \vec{S} \cdot \hat{n} dA$.} (spatially integrated radiant flux) spectrum of the probe return from the entire target's surface provides more information than the small spot size in that phase information over time is embedded in the sensed signal \cite{MSthesis:kobold06}.  On-average spectral elimination (or spectral reduction).\footnote{The laser vibrometry industry coins the term \emph{spectral elimination} to represent reductions to represent destruction or distortion of spectral ID features. Other industries use the same term for entirely different concepts.}

%Details follow in the body of this document.  The choice of sensing (small versus large spot size, imaging versus non-imaging) changes the quantity of spectral knowledge that the sensing system develops.

%%It is clear from the literature, simple analytical models, the FEA, and industrial experience that (1) the lower frequency modes have sharp resonances that are
%%
%%modes of nearby components through joints, including welds, which have nonlinear load-deflection curves. The energy at each of the modes in Figure 1 transfers to lower energy modes if physical load paths exist that are conducive to mode coupling.  Each

% \tableofcontents

[End of the table of contents.  This will be removed in the published paper.]

\subsection{Three observational characteristics}

\begin{figure}
\resizebox{14cm}{!}{
\includegraphics{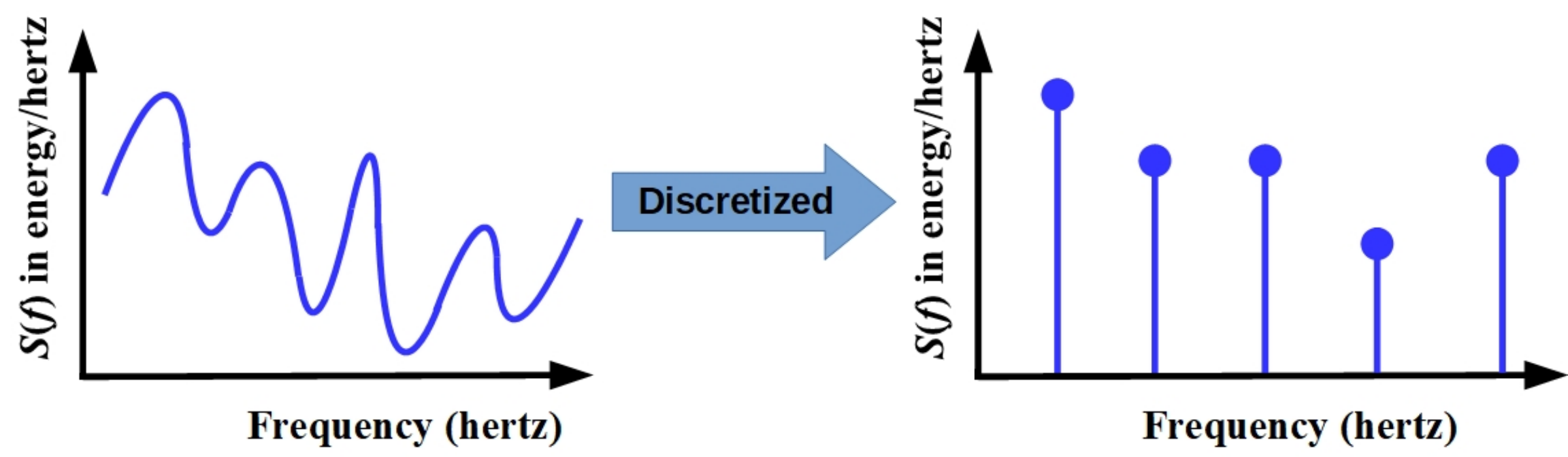}}
\caption[discreteModes]{\label{discreteModeSketch}The lowest frequency modes have high resonance quality \textit{Q} compared to high frequency response, unless systems have large damping on these ``lower" modes or frequency-dependent control systems. For adequate signal to noise ratios, these frequencies identify what are effectively discrete modes, as represented in the stem chart to the right.}
\end{figure}

\begin{figure}
\resizebox{8cm}{!}{
\includegraphics{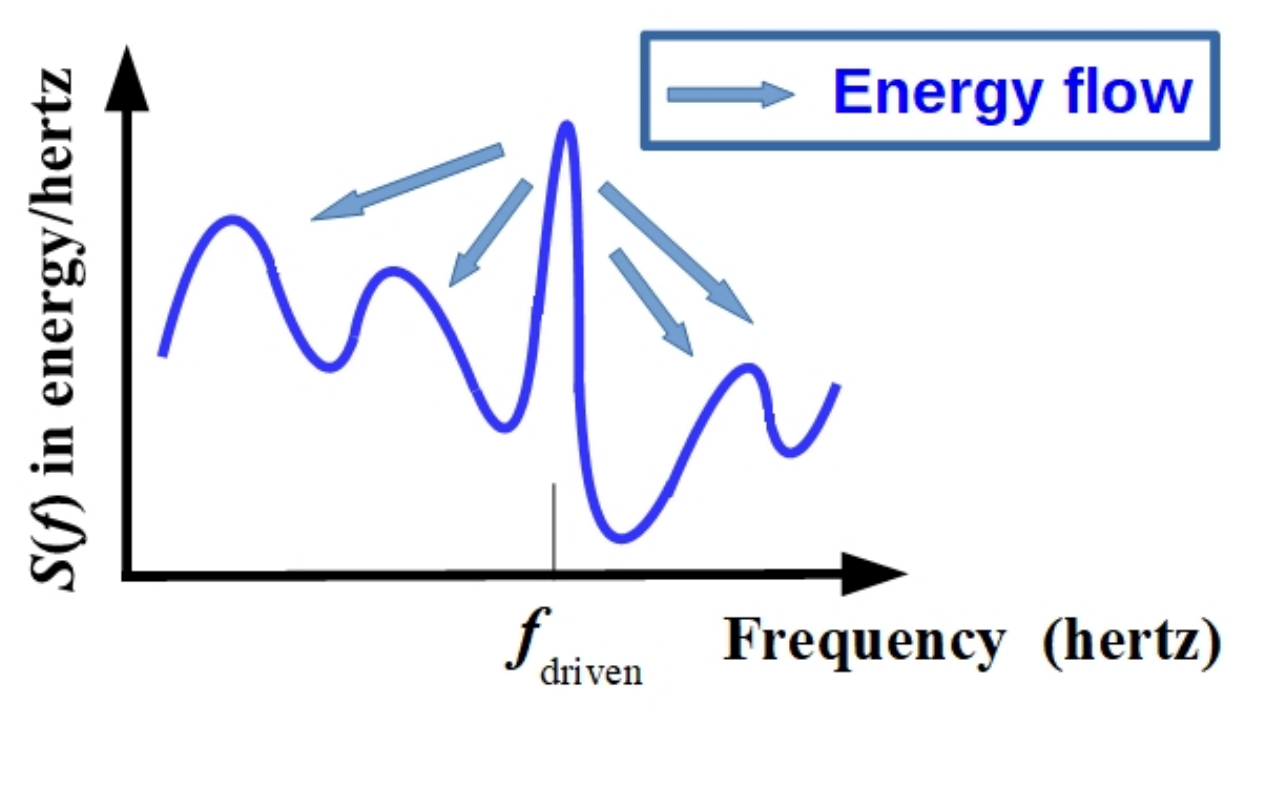}}
\caption[transferEnergy]{\label{energyXfrSketch}Vibrational strain-energy from different components transfers from a driven mode, the indicated most powerful response, to lower energy modes at different frequencies.  For example, the nonlinearity of all common, practical fasteners (bolted and riveted joints) allows energy transfer.  In this notional case, a high-\textit{Q} response at the driving frequency bleeds energy into other modes through friction or contact, a nonlinear process.  If the response at the driving frequency, $ f_{\mbox{\tiny{driven}}} $, has a wider full width at half height, then energy would transfer faster for nonlinear systems.  Some energy would then also transfer for linear systems as well, more so modes with a smaller frequency difference $ \Delta f = | f_i - f_{\mbox{\tiny{driven}}} |$.}
\end{figure}

\begin{figure}
\resizebox{8cm}{!}{
\includegraphics{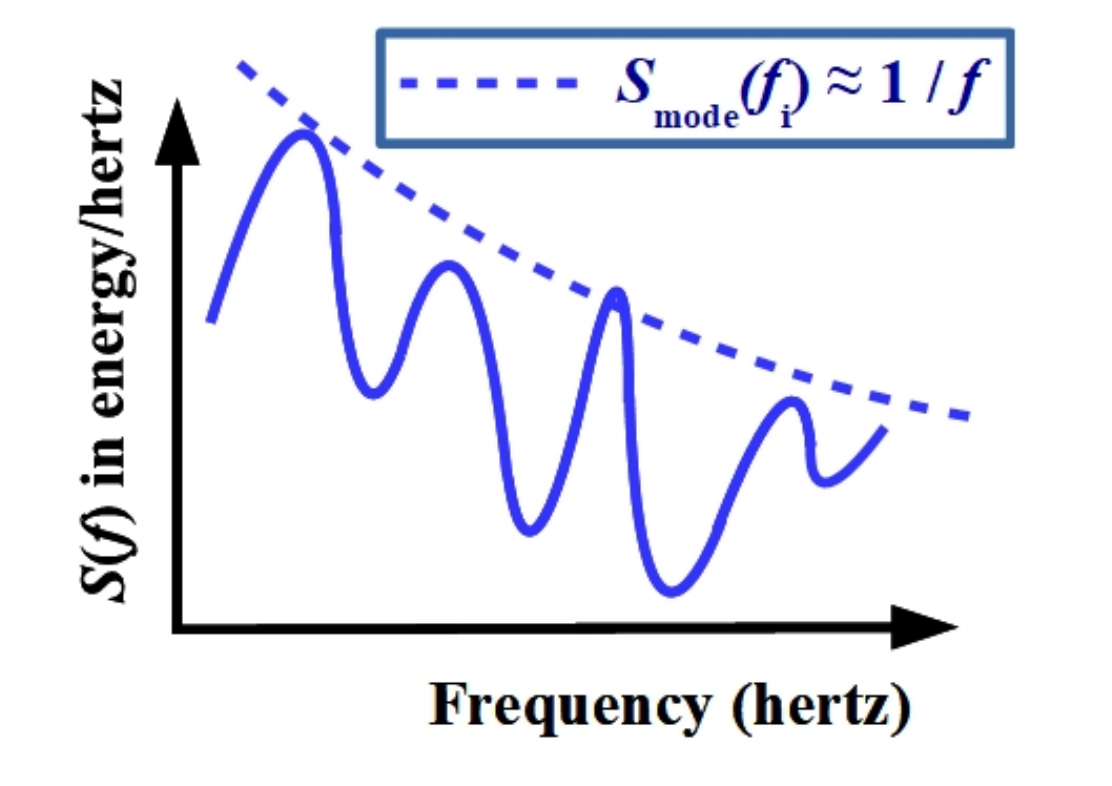}}
\caption[modalInverse]{\label{modalOrderSketch}As described by the Zienkiewicz quote in Section \ref{ZienkiewiczEnergyFlow}, high frequency modes eschew energy that then tends to flow into lower frequency modes.  These lower modes tend to ``fill up" first and they retain more energy than higher frequency modes.  There are at least two exceptions: the lack of vibration coherence can block the energy transfer, and interfaces that act like active systems can manipulate the energy or act like passive structural filters.}
\end{figure}

Based on the three previously mentioned supporting results, the experiments, simulation, and analytical results, three observations become apparent:  (1) the lower frequency modes have sharp resonances that are effectively discrete responses. Figure \ref{discreteModeSketch} is a sketch of a notional representation of this case.  These modes can be modeled on center frequencies when damping is limited to structural damping \cite{inProc:gByTedRose01}, as later discussed in Figures \ref{allModes} and \ref{missedMode}.  (2) energy transfers from one structural component and its vibration modes into the lower modes of nearby components through joints, including welds, which have nonlinear load-deflection curves.  Each mode is comprised of a 3-D mode shape (deflection vectors) of the entire structure (Figure \ref{energyXfrSketch}).  The strain-energy (modal energy) of each mode couples to lower energy modes as the system transforms from transient to steady state.  Finally, (3) modal energy flows into the lowest frequencies first, saturating them (Figure \ref{modalOrderSketch}), depleting the energy flow as the mode number increases from $f_0, f_1, f_2, \cdots$ and up through the closely-spaced high frequency modes that become practically infinitely dense.  For discrete modes the finite element analysis (FEA) results are limited to the number of degrees of freedom (DOF) in the model.

% \includegraphics{modObsrvCharAdiscreteModes}}

% modObsrvCharAdiscreteModes2

%\begin{figure}
%\resizebox{9cm}{!}{

The energy at each of the modes shown in the notional sketch (Figure \ref{discreteModeSketch}) transfers to lower energy modes if physical load paths exist that are conducive to mode coupling.  These sketches summarize decades of trade secret testing.  At lease one example appears in the literature for automotive modal analysis, ride and handling \cite{inProc:Pininfarina}.  Helicopter data is available from MIL-HDBK-810 \cite{man:MIL-HDBK-810}.  A transfer of energy from a strongly driven mode to other modes appears in Figure \ref{energyXfrSketch}.  That driver might be in a component located far from the receiving modes \cite{book:BendatNonlin98}, but in `frequency space' these modes might be coupled.  Vibrations might be driven by the engine at one frequency, yet sensed in the back seat at another frequency.  Nonlinearities in response can create harmonics of resonances that allow energy to couple with lower modes.  The low energy tails of wide resonances (damped broadening) also ``touch'' other modes of lower frequencies.  Another energy transfer mechanism occurs in clattering plates where spatial distribution of mode amplitudes causes contact that excites different modes. This transfer of energy is much similar to plucking a musical string at different locations to change the overall sound.  Energy transfer between states depends on the frequency of the mode, which is the square root of the eigenvalue in an analysis of normal modes.

% modObsrvBmodalParticipationChanges
%
%\begin{figure}
%\resizebox{6.5cm}{!}{

Figure \ref{modalOrderSketch} is a sketch where the final steady state of the vibrational system has energy ordered from strong resonances at the fundamental frequency to a paucity of energy at higher frequencies. A quote from Zienkiewicz, displayed later in Section \ref{ZienkiewiczEnergyFlow}, provides an explanation for this common 1/\textit{f} dependence of modal energy.  The exceptions noted also include strong damping or added mass.\footnote{In 1990 a team worked with Walker Automotive for General Motors vehicle Production Engineering.  Weights were placed along the tailpipes, mostly before the muffler, to quiet the vibration and acoustic noise -- except in the case of the Corvette (a design reversal in response to customer complaints that they had become \emph{too quiet}).  This was the testimony of one of the Walker engineers as a contrary example.  Tuning the sound for deep rumble has become well-known recently as a public selling point for many sports vehicles.}

% modObsrvCenergyEschewsHighFreq
%
%\begin{figure}
%\resizebox{6cm}{!}{

The models, structures, and observational characteristics that comprise this article are summarized in the Conclusion in Table \ref{TableI}.

\subsection{Physical model: symmetries, non-imaging, and coverage}

\begin{figure}
\resizebox{14cm}{!}{
\includegraphics{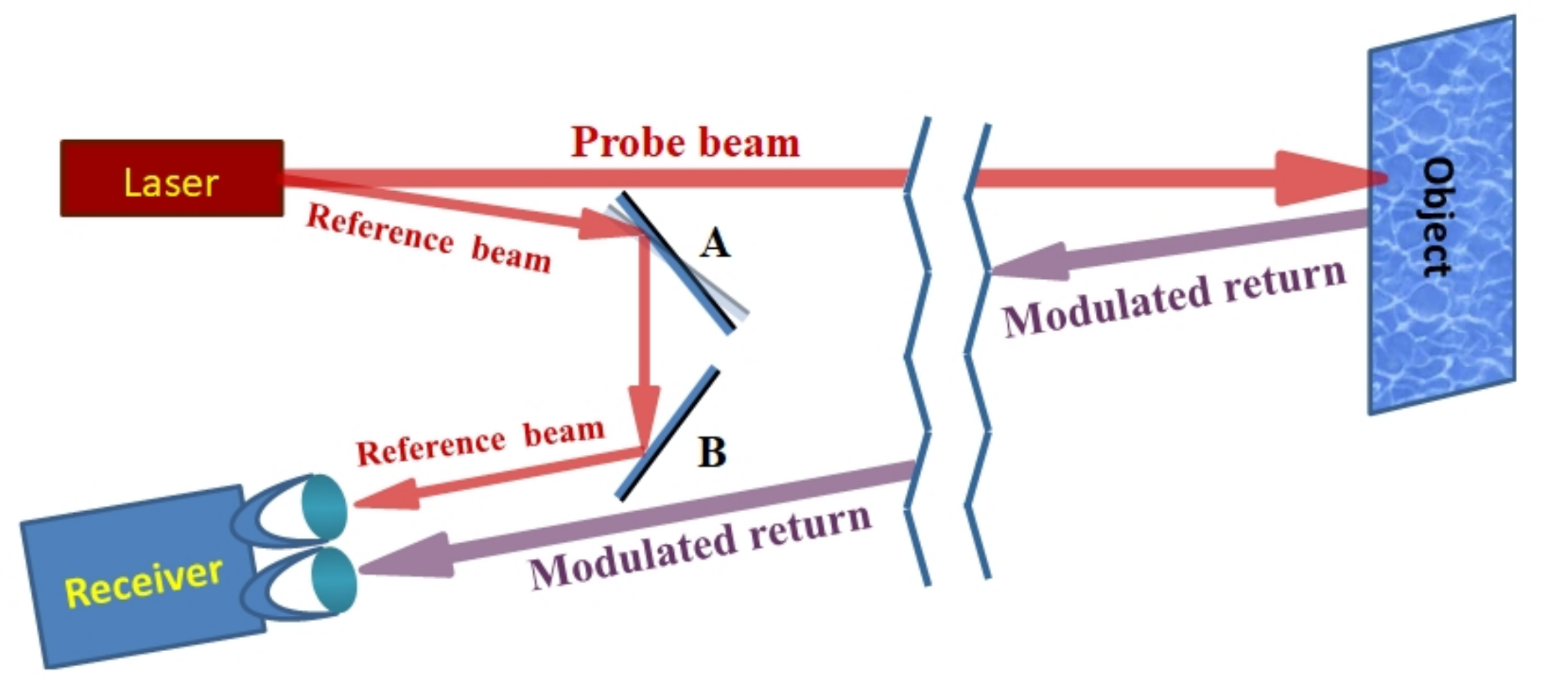}}
\caption[Laser Vibrometry]{\label{basicSkematic}Coherent radiance undergoes a spatially harmonic phase modulation as a function of the probe beam location due to target vehicle surface skin structural vibration.  The Object modulates radiance from the outbound probe beam at a distance of up to several kilometers while the receiver system keeps track of the phase of this exitance using the reference beam from articulating beam-splitter A and return mirror B, and appropriate delays (delay fiber or other systems not shown) where appropriate.}
\end{figure}

By using a ``pencil-thin" beam, conventional laser vibrometry systems can sometimes better identify modes in a simple academic structure such as the solitary vibrating bar in the following photos.  However, a large spot size can adequately measure those modes and more.  This can be shown by simulation and analytical calculation. The simulation uses FEA for vibration deformation data that is handed-off to a set of MATLAB\texttrademark functions that perform Fresnel propagation for the optical sensing. The FEA shows the mode shapes and calculates the energy per mode which are the modal participation factors (MPFs) \cite{inProc:MPFbyTedRose98, book:Thomson81}.  The FEA surface vibration results are input for the MATLAB image propagation calculation (Fresnel propagation \cite{book:GoodmanFO68}).  The physical schematic for the simulation appears in Figure \ref{basicSkematic}.  Laser vibrometers use many different methods that can have an effect on the precision of the measurements.\footnote{This interferometric model allows analysis of the phase modulation of the exitance from the vehicle (the \emph{return}), hence its use for the optical analysis.  At some level, even the laser vibrometers that use Doppler to measure vibration have images that are affected by phase modulation generated by vibration mode shapes.  This is part of the mechanism of spectral elimination.  Phase modulation is used here to determine the effect of mode shapes on the return for both imaging and non-imaging systems.}

% \includegraphics{opticalModel4Return.eps}}

%\begin{figure}[LzVibConcept]
%
%\includegraphics{opticalModel4ReturnShort}}
%\caption[Laser Vibrometry]{\label{basicSkematic}Coherent radiance

%\begin{figure}
%\resizebox{8.4cm}{!}{

The laboratory measurements of the vibrating bar show the signal to noise ratio (SNR) is quite high for structural components and, unless the system is purposely damped, the resonances are sharp.  These sharp resonances lead to the discrete mode picture (stem plot at right in Figure \ref{discreteModeSketch}) that the field of modal engineering evolved to exploit.  The literature provides the theoretical reasons that strain-energy tends to eschew higher frequency modes and thus populate the lower modes.  Years of experience of vehicle design and analysis support these conclusions, as does the FEA of clattering armor plate reported here and in detail in a thesis \cite{MSthesis:kobold06}.  The FEA results show that unsymmetrical modes\footnote{\label{symmNote} The one-dimensional model in the Appendix uses symmetric and antisymmetric 1-D modes.  This solution applies to the symmetrical and unsymmetrical modes resulting from the FEA for the tank hull and armor.  The latter do not have full symmetry because the surfaces are never perfectly flat.  Symmetrical bending `$ \approx $' maintains the gap the same across the surface, while unsymmetrical `$ \simeq $' has both compressive and open gaps at different locations at the same time with antisymmetrical (not mentioned) being alternately open and closed `$ \asymp $.'  In 3-D the average centroid surface is almost never perfectly flat and so a symmetrical mode cannot be geometrically symmetric, but a 1-D model of it can.  See \cite[p 31 fig 3]{MSthesis:kobold06}.} of the clattering plates have higher frequencies than symmetrical (in-phase) vibration of the parallel plates, and that these low frequency modes are discrete below 500 Hz, as is the case with typical automobiles.  These modes are usually easy for laser vibrometers to detect.  The optical image propagation model shows that even a non-imaging spectrum of the surface of the clattering plates contains energy-ordered modes whose vibration strain-energy drops off with frequency, as described in the Zienkiewicz quotation provided here in Section \ref{ZienkiewiczEnergyFlow}.

Commercial laser vibrometers tend to use pencil-thin spot sizes on the object being measured.  This spot's cross-section can be as small as a one pixel response.  Large spot size \emph{imaging} systems have far more information, much of which can help spectral identification (ID), such as the method developed herein.  Further, \emph{non}-imaging signals average responses from all pixels for each time step.  An optimum automated target recognition (ATR) system could use a combination of small spot size and full size illumination.  But a large spot size tends to help tracking in the paint-the-target phase, while still enhancing spectral ID.

\section{Structural nonlinearities}

\label{structNonlinSection}

Several different types of laser vibrometers can measure dynamical properties of vibrating surfaces.  The type of laser vibrometer is less important than the phenomenology of what can be measured remotely.  A priori project restraints such as requirements to calculate a transfer function, or to use pencil-thin probe beams, can limit the sensed spectra and thus omit or change perceived vehicle behavior(s) in a manner unrelated to errors in measurement or analysis.  In many fields such as acoustics, the modal results can be less accurate than a kinematic approach would provide.  Nevertheless, especially for complicated nonlinear systems, it sometimes pays to ignore detailed kinematic analysis, which are often invalid for nonlinear systems, in favor of \emph{modal engineering} to tabulate the modes (deformed shapes for particular resonance frequencies), their frequencies, and their energy or \emph{modal participation factors} \cite{book:Thomson81}.

%The comparison of two forms of this remote sensing shows one aspect of the information missing in some typical spectral analyses subject to keep-it-simple project constraints, such as demanding dynamical transfer functions or using pencil-thin beams to probe the surfacel

%%Modal engineers have routinely done nonlinear analysis for automotive and other vehicle systems for over half a century by concentrating on identification of the modes and their frequencies as opposed to using transfer function (because that would require full linearity).  This work uses a structural model of a plate bolted to another plate on a base structure.  The work also uses simplified and even ``linear" models for some components, but they cannot deliver the behavioral metrics that the full model

For over half a century modal engineers routinely used modal analysis for nonlinear vehicle structures.  This work uses a structural model of a plate bolted to another plate on a base structure.  The work also uses simplified and even ``linear" component studies.  These studies cannot deliver the behavioral metrics that the full system model provides.  Component studies do verify limited aspects of the full nonlinear model, which is a crucial part of the overall analysis.  In the full analysis, the structure is a free-floating system using D'Alembert reactions, an analysis technique common to vehicle engineering.  See, for example, the numerous `quarter-panel models' in the literature \cite{ppr:mitra06,misc:koboldExperience}, which are mostly meant to show \emph{how} to model some features for reasons stated previously.\footnote{As of 21 November 2019 only the first two hits for ``quarter-panel dynamic response'' were related to vehicles. No experimental results showed up in the top (most pertinent) 10 pages of 10 hits each of 1,050 hits.}

The sensor model is an optical model without the usual air turbulence analysis, except to check for Fante's wavefront coherence breakup range \cite{ppr:breakupFante75} and a few other imaging issues \cite{MSthesis:kobold06} that are out of the scope of this paper.

The physical model for the clattering armor assumes a linear superposition of vibration modes on a nonlinear structure.  This is not a linear time-invariant (LTI) system.  The details of removal of these typical LTI assumptions and its relationship to stability, stabilization, and stabilizability, are in the thesis \cite{MSthesis:kobold06}, available through DTIC dot mil. The analysis includes a treatment of how changes in fixity causes changes in the modes. Similar changes can model the aging of parts.  Therefore, the modulation of the sensed optical radiant flux is related to the analytical nonlinear expressions in the Appendix that help explain the contact nonlinearity in the FEA simulation.  The modeling complications related to fixity, nonlinearities, and modal participation factors \cite{inProc:MPFbyTedRose98} are dependent on control of the nonlinear solution per time step and over time using load increments, iterations, and other parameters.  The analyst also needs to make sure the solution follows the millions of load deflection curves \cite{book:Zienkiewicz91, MSthesis:kobold06} in the full system model.

The finite element analysis (FEA) output represents the three-dimensional structural model of the clattering armor plate system and provides the time history of a structural vibration that modulates the optical return recorded by the sensor.  The Appendix contains a set of single and two degree of freedom analyses that result in analytical expressions.  The objective of the analyses is to use the optical images, or non-imaging radiant flux (watts), to determine the frequency of structural spectral modes for target identification (ID).

%result in analytical expressions that provide insight into the FEA results.  Discussions concerning the need to have load increments, iterations, and other aspects of the FEA solution follow the millions of degrees of Freedom (DOFs), each with their own load deflection curve, are found in the author's thesis \cite{MSthesis:kobold06}.  The objective of the analyses is to use the optical images to determine the frequency of structural spectral modes for target identification (ID).

\subsection{Modal Participation Factors}

Recent conversations with Navy scientists and engineers tend to evolve into studies of the role that the nature of system observation plays in the perception of the modal system, and how the low frequency strain-energy modes of the structure are effectively quantized.  The discrete nature of the modes has been an underlying assumption of modal analysis and in its application to ride-and-handling for half a century.  The modal participation factors \cite{book:Thomson81} describe the dynamics of energy flow by using vectors of MPF's that span the pertinent modes \cite{inProc:MPFbyTedRose98}.  Each MPF is an element in vectors $ \mathbf{\Phi[\mathit{f_i}]} $ that are arrays ordered by mode number \textit{i}, used in the FEA of vibrating structures such as vehicles, antennas, and spacecraft.  For example, NASTRAN\texttrademark may require `DMAP alters' (macros) to read out some of the components of $ \mathbf{\Phi[\mathit{f_i}]} $.

An example of the unintended consequences of design decisions for the 1990 era Corvette follows this detailed restatement of the three modal characteristics introduced earlier, which are essential to the correct system ID.  They will be used in the example.  (1) The lower frequency modal states (eigenvectors) form discrete modes, even for the complicated system synthesis models and the vehicles they represent.  The work of vehicle design for these issues focuses on (2) transition probabilities\footnote{Transition probabilities are related to structural coherence spectra.  Modal engineers use them to reduce undesired vibration modes (resonances) in the structure.\cite{ppr:AllemangMAC03,phdth:Allemang80}.} and (3) the flow of strain-energy that preferentially fills the lower modes with energy, as discussed in the quoted passage in the next section.

\subsection{Observational characteristics: Discrete, Transition, and Ordering}

1. \emph{Discrete modes} are countable sets of distinct and high signal to noise ratio (SNR) response of bandwidth narrow enough to be obviously a single mode.  The strain-energy levels within each modal state are tabulated by the structural design team in order to determine critical components.  These critical regions absorb most of the ride-and-handling and structural engineering effort during vehicle design, resulting in a final list of lower frequency energy levels\footnote{Acceleration spectra are measured in units-per-hertz such as strain (strain-energy amplitudes) or acceleration.  In practice the ``power" is displayed, signal processing ``power'' being the square, element-by-element, in $g^2$/Hz.  However, the optically sensed units are dB/Hz, which for non-imaging is re radiant flux $\Phi_{e,ref}$ in W/Hz).}  The lower modes relate to modes shapes that are usually characteristic of the structure, such as bending, twisting, and stretching.  A real-life example appears later in this section.

2. \emph{Transitions between modal states}, from one eigenvector of the structure to another, are often sigmoid in nature\footnote{While there are systems where joints change behavior on a short-term basis during operation (e.g., magneto-rheological and similar semi-active systems), the sigmoid transition is usually driven by two factors: (A) energy thresholds (e.g., bolted joints have a bi-linear load curve \cite{man:aeroStrNASA}), and (B) design changes during the drawing release phase of vehicle development.} where the region of rapid mode coupling depends on meeting strain-energy thresholds that allow more efficient coupling between modes.  Most of the energy tends to flow to components with lower fundamental modes according to characteristic (3) below.  Nearly all joints are structurally nonlinear\footnote{Even spot welds have at least geometric nonlinearities in the Green's strain tensor. This is due to large strain related to dissimilar stiffness and offset load moments (`kick's' in aerospace jargon).  Nonlinearities are usually due to frictional hold of bolted or riveted joints that ``\emph{must be ignored}" in most structural analyses \cite{book:Popov98}.  Therefore, most system FEA cannot model MPFs unless the model is nonlinear and designed to develop MPFs.} for the transfer of forces and reactions, a situation that provides substantial coupling nonlinearity.

3. \emph{Ordering of modal participation:} The low frequency end of the spectrum tends to hoard the majority of the strain-energy, as the quote from Zienkiewicz explains.  These characteristics are summarized in the table \ref{TableI}, the Conclusion.  Analysis of MPF's (developed by characteristic 2) can determine state transition probabilities.  However, modal engineers tend to be focused on solving critical failure issues.  Engineers might use similar terms for entirely different kinds of `modes' such as `failure mode effects analysis' (FMEA).  However, a large part of modal engineering involves creation of a record of the order of the vibration modes, followed by an analysis of the tracks of the modes as the design variable change.

%%Getting all the parameters of a new vehicle to work together is a work of art, no matter how many thousand Ph.D. level engineers are working on it -- and General Motors alone uses nearly a thousand of them.  Even technical engineers and scientists who do not work a Production Engineering department are surprised to learn how close new designs come to failing, and the number of compromises that are necessary to get the entire system to work.  From a statistics point of view, it is a miracle that automobiles work as well as they do.

A structural vibration example described below shows how a structural design change can de-couple modes in order to improve ride-and-handling.  However, this change has consequences other than aesthetic design constraints.  The changes can increase crashworthiness risk and other seemingly unrelated systems, and they can degrade previously adequate vibration and modal engineering balances in the design.  For example, stiffening one joint or component usually provides a much more efficient flow of energy through that stiffened element of the structure.  Strain-energy is drawn from the high frequency modes to the low modes in a surprisingly efficient manner once such a structural ``channel" is open (components have structural coherence).  Then the MPFs start to re-balancing as the entire integrated structure starts to equilibrate.  So a stiffener can solve one problem and cause another.  The engineering group that deals with braking and crash loads that transfer from the front to rear bumper can have its margins go from adequate to negative due to a design change by the ride-and-handling team that solves a mode-coupling problem.  This is similar to an old solution to convertible mode coupling:

\subsection{Inter-organizational effects of design changes}

The 1980 era Chevrolet Corvette convertibles an initial design had a fundamental bending mode just below 30 Hz, which put it very close to the suspension mode.  This is a overall vehicle bending mode that curves about a lateral axis, where the bumpers move vertically and together, both in direct opposition to the vertical motion of the seats. It is usually the lowest frequency vehicle mode for a convertible (as if the vehicle is trying to do sit-ups).  The Corvette designers needed to stiffen this mode to reduce coupling to the nearby suspension modes, which were also near 30 Hz. The models in the late 1980's had very tall rockers that satisfied this requirement.  This was the most efficient structural fix because the stiffness of a beam to bending vertically is proportional to the cube of its section height \cite{book:Roark89}.  Apart from the aesthetic issue with having to step high to get into the Corvette, these tall rockers also changed other load paths including a for-aft transmission of load, important to crashworthiness, as well as making the vehicle slightly more heavy.  Automotive is one of the few industries that has a price on the engineering sufficient to remove a kilogram of mass, which was \$50,000 in 1990.  These and other costs were accepted in order to move the fundamental bending mode away from the suspension mode to effectively eliminate coupling.  In later years, other technologies helped solve this issue.

With the exception of the Pininfarina chassis test results \cite{inProc:Pininfarina}, few modal analyses such as these (chassis or body-in-white) are in the literature.  To a scientist, the ride-and-handling issues and how they relate to the three observational characteristics described above might seem to be basic enough to warrant several scholarly articles.\footnote{Physicists in Ann Arbor. mostly graduate students in 1998, were baffled, surprised that a wheelbase change of less than 2 cm would require over a year of mechanisms analysis to re-balance just the toe, caster, and camber - in negotiations with competing requirements from groups that engineer tires, turn radius, vibration, and crashworthiness.  Most vehicle designs barely consolidate the competing requirements, which is more difficult than it would seem.}  However, this is clearly the arena of trade secrecy at such a high level of value that corporate lawyers would be unsurprised at the paucity of measured data available to the public.

\subsection{Measured data: Vibrating clamped-clamped bar}
\label{sect:measuredVibClamped$^2$}

\begin{figure}
\resizebox{9cm}{!}{
\includegraphics{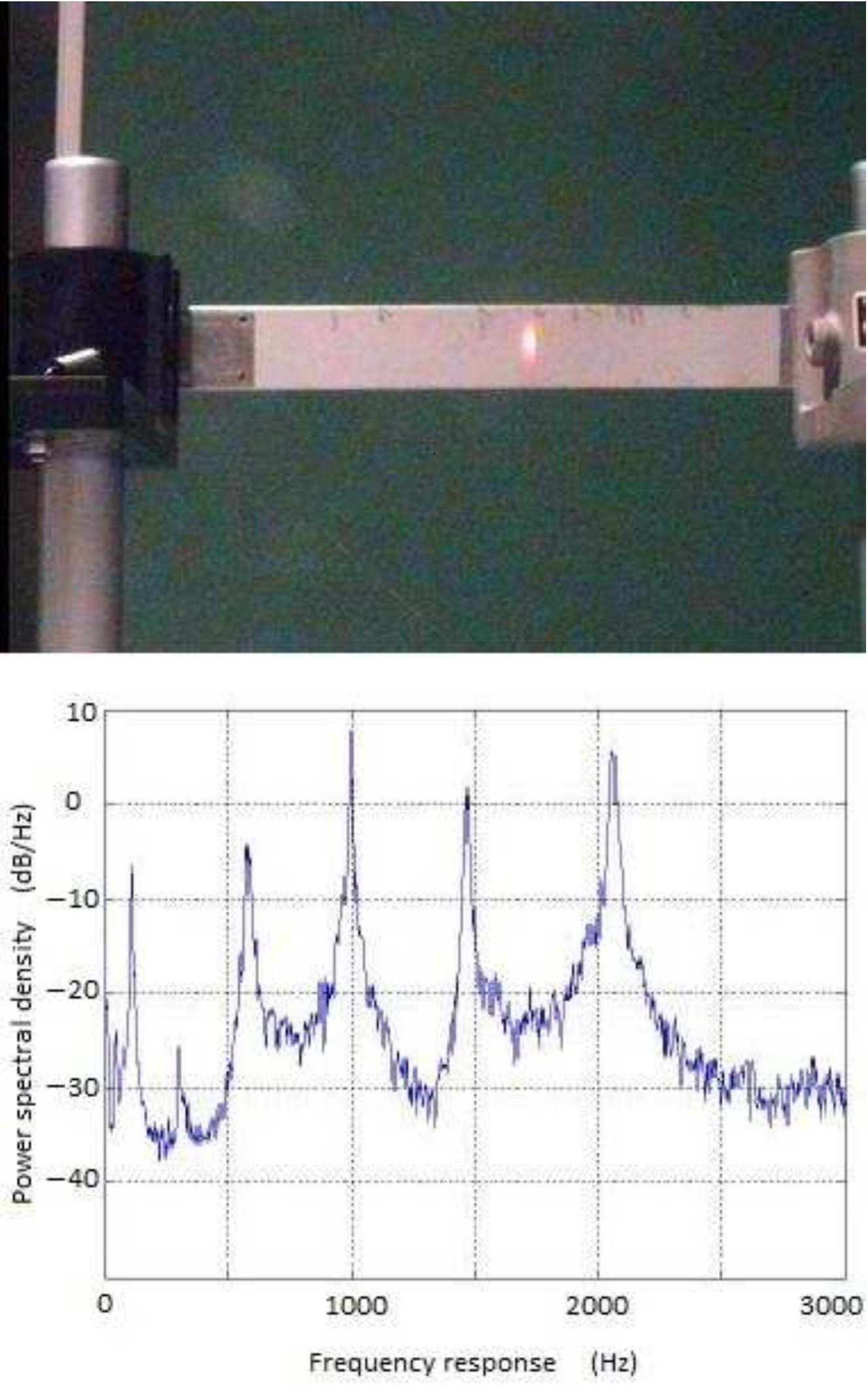}}
\caption[Spectral Elimination]{\label{allModes} The image from a small spot size makes a spectrum with a $f_3$ = 1460 Hz mode. Permission N. Pepela \cite{MSthesis:NgoyaPepelaLaserVib03,ppr:insensitive}.}
\end{figure}

\begin{figure}
\resizebox{9cm}{!}{
\includegraphics{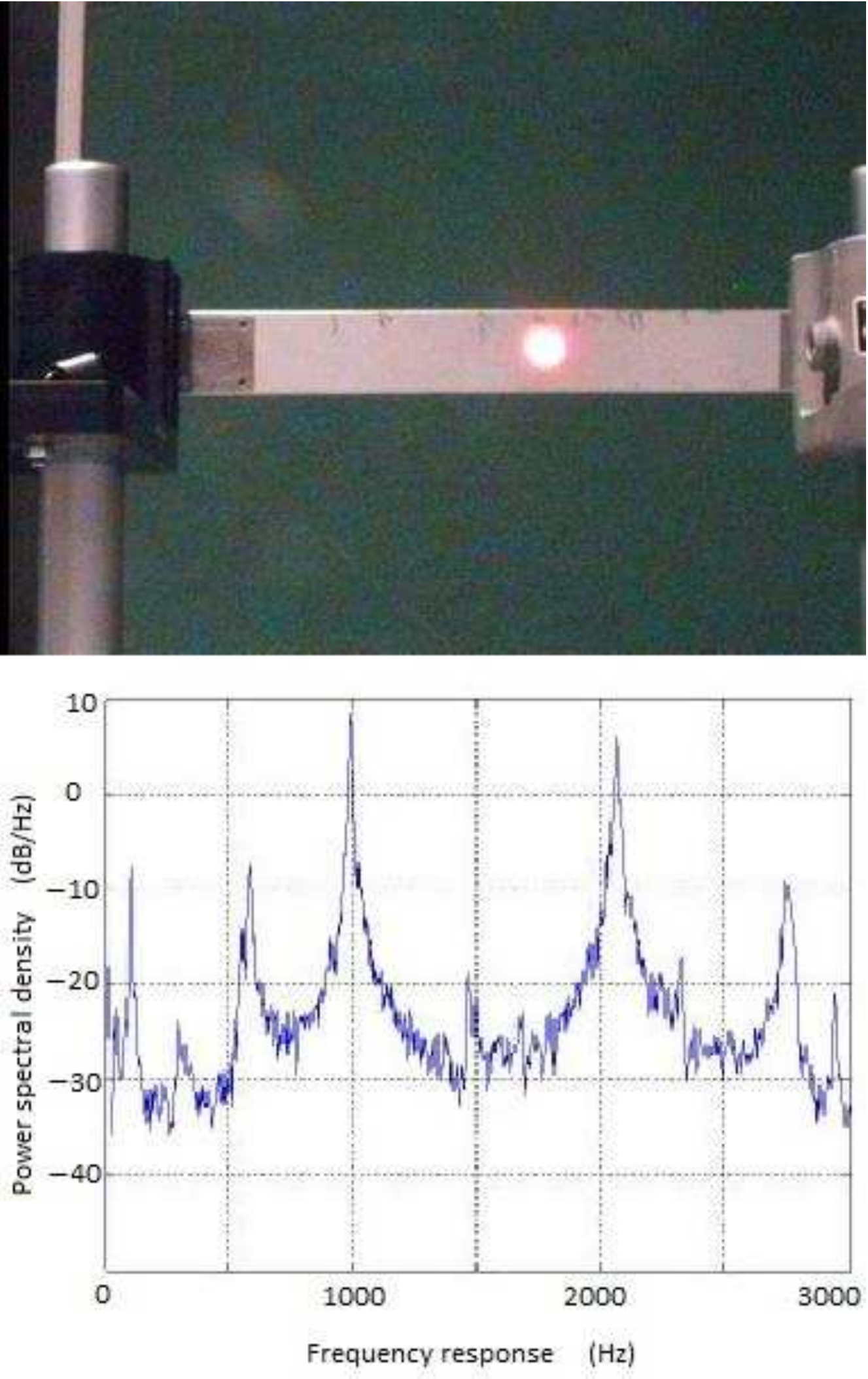}}
\caption[Spectral Elimination2]{\label{missedMode}A larger laser spot size creates a laser vibrometry spectrum where the $f_3$ = 1460 Hz mode phase-averages away.}
\end{figure}

In his thesis Fl. LT. Ngoya Pepela (he was an Australian Flight Lieutenant in 2006) showed that the modes are  not just maxima of a spectrum, but spikes with huge SNRs in the spectral response \cite{MSthesis:NgoyaPepelaLaserVib03}.  These are the ``discrete'' modes described above in `Observation 1.'  His Figures \ref{allModes} and \ref{missedMode} each show a photo of the vibrating bar illuminated by a laser vibrometer, with the corresponding spectral density plot are beneath the photos, plotted in the lower pane of each figure.  The first (Figure \ref{allModes}) shows all the major modes whereas the second (Figure \ref{missedMode}) is missing the modes at 1460 Hz.  In Figure \ref{missedMode} the vibrating bar is seen to be illuminated by a laser beam with a large round spot size centered on the 1460 Hz node of the vibrating bar.  In Figure \ref{allModes} the right half of the laser beam of Figure \ref{missedMode} is blocked, illuminating only the left half of the symmetric 1460 Hz mode shape.  The halved beam size (Figure \ref{allModes}) allows the laser vibrometry spectral analysis system to show a stronger 1460 Hz mode return due to the elimination of spatial averaging.  That averaging is introduce by phase-related destructive interference from both sides of the 1460 Hz mode shape in the full spot size result of Figure \ref{allModes}.

%are shown below (permission granted in a prior article \cite{insensitive}.  The lab photo of the illuminated vibrating bar is shown in the upper pane with its laser vibrometry response shown below in the same figure.  These plots are shown in two separate cases, one with a large (round) spot size versus the case with a reduced (blocked) spot size.  The small spot size (Figure \ref{allModes}) allows the laser vibrometry spectral analysis system to show all the modes without spatial averaging that might introduce phase-related destructive interference.
%
%\begin{figure}
%\resizebox{8.4cm}{!}{

% SpectralElimNgoyaPepelaFig2smallSpot

Initially spectral elimination might appear to be a problem for the laser vibrometry industry for use with non-imaging sensors if the spot size is large.  However, Ngoya Pepela's thesis \cite{MSthesis:NgoyaPepelaLaserVib03} and another article by the author \cite{ppr:insensitive}, with support from simulation and analytical calculations, show numerically how unlikely spectral elimination is for commonly manufactured items, even if the spot size encompasses the entire vehicle.  Furthermore it is necessary, but not sufficient that super-symmetric structures produce spectral elimination in laser vibrometry \cite{MSthesis:kobold06}.  The vibrating bar is the simplest form of structural super-symmetry that results in spectral elimination, but it is rarely the only structural component a laser vibrometer can use for identification of a manufactured structural system (vehicle), unless the target is the vibrating bar (a structural \emph{oscillator}) that requires identification.  But even for such oscillators, the typical transducer is a piezoelectric system that is not super-symmetric, and thus not prone to spectral elimination; the source of oscillator energy is separate from the driven bar.  Therefore, a vehicle that is purposefully fitted with vibrating bars can still be identified with its spectral ``fingerprint" in spite of possible spectral elimination from oscillators.  Pepela's result \cite{MSthesis:NgoyaPepelaLaserVib03} was that use of small spot sizes reduces spectral elimination from the laser vibrometry spectral result (Figure \ref{missedMode} vice \ref{allModes}).

%\begin{figure}
%\resizebox{8.4cm}{!}{

%  SpectralNonElimNgoyaPepelaLargeSpot

\subsection{The McKinley observation}

In early 2018 Michael McKinley of Arlington, Texas, observed that Figure \ref{allModes} contains modes (at approximately 2740 and 2900 hertz) that are not visible in the small spot size collection \cite{misc:emailFromMcKinley} in Figure \ref{missedMode}.  While the small spot size is still large enough to average out higher frequency (smaller vibration shape wavelength) modes, another spectral elimination that small spot size vibrometers are susceptible to is the effect of the probe spot being on a Chladni line of nodes \cite{book:RayleighTOS1} of the vibration shape for particular frequencies \cite{book:ChladniWiki}.  Detection of these modes at 2300, 2740, and 2900 Hz is a positive observational difference provided by using a large spot size, in this case.

Therefore, spectral reduction and elimination (SR and SE) happens for all spot sizes, large and small, for super-symmetric structures such as 1-D bars.  The prior paragraph discusses why this is not a problem for manufactured vehicles; nature provides a mitigation (vibrometry is a useful classifier system for realizable structures) of the perceived problem that the McKinley observation clarifies (small spot sizes do not mitigate SE).

\subsection{Observational differences}

%%The difference between small spot size (Figure \ref{allModes}, and larger spot size (Figure \ref{missedMode}), and full illumination (Figure \ref{basicSkematic} and \ref{impactResponseFRF}) comprises one dimension of observation variation that can

%The difference between small spot size (Figure \ref{allModes}, and larger spot size (Figure \ref{missedMode}) comprises one dimension of observation variation that can change the observer's view of reality.  Several physical systems such as this automotive design example can be used to explain the existence of the three characteristics of modal systems listed above.  It is the contact nonlinearity that is the main physical model of interest in this article because contact is a relatively simple form of nonlinearity.

The difference between small spot size (Figure \ref{allModes}), and larger spot size (Figure \ref{missedMode}) comprises one dimension of observation variation that can change the observer's view of reality.  Physical systems such as the Corvette design example can be used to explain the existence of the three characteristics of modal systems listed above.  Contact nonlinearity is the main physical model of interest in this paper because contact is a relatively simple and ubiquitous form of nonlinearity, found in nearly all structural joints.

\section{Energy ordering}

\label{ZienkiewiczEnergyFlow}

Vibration strain-energy transfers from one component to another, usually through joints that are necessarily nonlinear (even spot welds), depending on the amount of `fixity' of that interface or joint.  An extensive treatment of how fixity changes the modes and therefore the sensed optical radiant flux is in the author's thesis \cite[B.2.3, p. 187]{MSthesis:kobold06}, including nonlinear examples that compare to the contact nonlinearity of the FE model.  Fixity is an engineering variable for a particular DOF that determines the ratio from zero to one (0\% to 100\%) of force, moment, or torsion \cite{book:Roark89} that will transfer from one component to another through the joint member for whom the fixity is defined \cite{book:Popov98}.

Even ``simple" vehicle structures have complicated load paths that are nonlinear (mostly through joints) where forces from one region of the vehicle affect parts elsewhere on the vehicle.  These are the pathways for vibration energy flow that tend to dump energy into low frequency structural modes.  Modal engineers use isolators and suppressors to stymie this natural tendency, but mostly just for critical components. Some components tend to have non-negligible vibration spectral energy down at these ``fundamental" and near fundamental modes\footnote{The `fundamental' mode is the mode that has the lowest frequency, contrary to the definition for a musical fundamental frequency.} even without the nonlinear effects. Damping helps drive the energy into the lowest modes because the resonance spreads (widens) to encompass a larger bandwidth.

Zienkiewicz showed energy transfer using only viscous damping \cite[340-341]{book:Zienkiewicz91} to form a ratio of damping to its critical value,\footnote{Critically damped systems remove all oscillation.} $c_i = 2 \omega_i c_i' $ depending on the modal frequency $ \omega / 2\pi $, as quoted below.  For vibration systems that are driven by a forcing function \textit{f}, Zienkiewicz describes the differential equation (DE)  $ \mathbf{M\ddot{a} + C\dot{a} + Ka + f = 0} $ using a mass matrix ($\mathbf{M}$), a damping matrix ($\mathbf{C}$), and a stiffness matrix ($\mathbf{K}$), acting on deflection and force vectors $ \vec{a} $ and $ \vec{f} $.  $ \alpha $ and $ \beta $ are damping parameters.

\textbf{Zienkiewicz quote:}

\begin{quotation} \label{quote:Zienkiewicz}
`` ... we have indicated that the damping matrix is often
assumed as $ \mathbf{C} = \alpha \mathbf{M} + \beta \mathbf{K} $.
Indeed a form of this type is necessary for the use of modal
decomposition, although other generalizations are possible
[references given in the book].  From the definition of $ c_i' $,
the critical damping ratio [described above], we see that this can
now be written as

$ c_i' = \frac{1}{2\omega_i}\mathbf{\overline{a}}_i^T
(\alpha \mathbf{M} + \beta \mathbf{K}) \mathbf{\overline{a}}_i =
\frac{1}{2\omega_i}(\alpha + \beta\omega_i^2)
$

Thus if the coefficient $ \beta $ is of larger importance, as is
the case with most structural damping, $ c_i' $ grows with  $
\omega_i $ and at high frequency an over-damped condition will
arise.  This is indeed fortunate as, in general, an infinite
number of high frequencies exist which are not modeled by any
finite element discretizations." \cite[340, 341]{book:Zienkiewicz91}
\end{quotation}

Therefore, vibrational strain-energy tends to naturally migrate from higher to lower frequency modes.  Some of the strain-energy can find its way down to the lower modes because over-damping causes more overlap of modal response resonances.  Nonlinearities add to the methods of energy transport as discussed later.

This results in a set of modes where the mode number increases monotonically with frequency and has an energy that decreases monotonically with mode number.  Possible exceptions to this natural effect include artificial structures\footnote{Examples of artificial exceptions to the 1/\textit{f} phenomenon could be exotic.  Perhaps a feedback system might ``manually" excite high harmonics without exciting lower harmonics, such tapping very near the base on a taut chord would tend to excite the higher frequency mode.} or temporary transients such as might be introduced with magneto-rheological fluid or other systems controlled by magnetic or electric field variations.  The 1/\textit{f} phenomenon is found in many different fields of engineering.\footnote{One example is the spectrum of a surrogate missile plume.  The power spectral density at low optical frequencies is large, decaying exponentially with higher frequency in an approximately straight line for a log-log plot.}

The local mode structure in Ngoya Pepela's laboratory test measurement in Figures \ref{allModes} and \ref{missedMode}, which do not follow the 1/\textit{f} phenomenon within the displayed 3 kHz,  \emph{appears} to be a contrary example. This an example of not being able to `see the trees for the forest' within its part of the response spectrum of the very stiff clamped-clamped structure.  For those results, the vibrating bar only had a half-dozen distinct modes across 0-3000 Hz, one of which was diminished in Figure \ref{missedMode}.  The utility of the structurally super-symmetric bar is that it shows discrete modes, and that the energy in some modes are spectrally reduced or eliminated by spatial ``averaging" -- integration over the area of a surface of variable phase that introduces varying levels of continuous interference depending in large part on the limits of spatial integration \cite{ppr:insensitive}.

%spatial integration \cite{ppr:insensitive}.  But is this really a counter-example of the 1/\textit{f} phenomenon?

That the vibrating bar of  Figures \ref{allModes} and \ref{missedMode} is consistent with modal ordering becomes apparent by analysis of the physics of a simple model of a point mass at the end of a spring of stiffness \textit{k}.  This explanation considers multi-variate non-uniform parametric change in the system.  The maximum potential energy is approximately $U_{\mbox{\scriptsize{max}}} \approx kx^2/2$ where \textit{x} is the largest extension of a simple spring.  Assume that the frequency $ f^2 = k/(4 \pi^2 m)$ is increased, then the energy grows as the square of the frequency, parabolically as $U \approx 2 \pi^2 mf^2x^2$.  In practice, the deflection \textit{x} decreases with frequency, as it must.  However, there is a practical limit to reduction in deflection.  When that limit is reached, while the frequency continues to increase, the potential energy must grow.  This means that higher frequency modes require more energy, even for smaller deflection -- especially when deflection is already small, when $x^2 \ll U / (2 \pi^2 mf^2)$.  The kinetic energy side of this simple calculation is \emph{Rayleigh's method} \cite{incoll:MarksHdbk79}.

It may help to see this from a force point of view using the gradient of $U$.  If the restoring force is conservative, it is $F = \partial U / \partial x = 4 \pi^2 mf^2x$, and is thus a quadratic function of the frequency, $f$.  It might be tempting to consider a ``conjugate force" for frequency, where the gradient $\partial U / \partial f = 4 \pi^2 mx^2f $ (units of action : Energy $\times$ time, or momentum $\times$ distance) \cite{book:CRChdbkChPh}.  However, the force ($F = \partial U / \partial x$), cannot increase ad infinitum with $f$.  At some point assumptions of linearity and structural integrity start breaking down.  Combining the observations of this upper limit on the force, and the understanding that the spring extension is limited to $x > 0$, the energy required to have a particular mode increases approximately quadratically with $f$.  This increase strain energy based on the modal displacement $ u^{\prime}_i \approx \int dx dy w(x, y) $ is ``on average" energy in this sense -- that the other variables are limited, and using experience that dynamical systems naturally tend to avoid populating the higher modes with energy.  Many children experience this effect when they try to forcefully excite a higher frequency mode in a rope, only to fail unless they exert considerable effort to the point of overexertion.

\begin{figure}
\resizebox{15cm}{!}{
\includegraphics{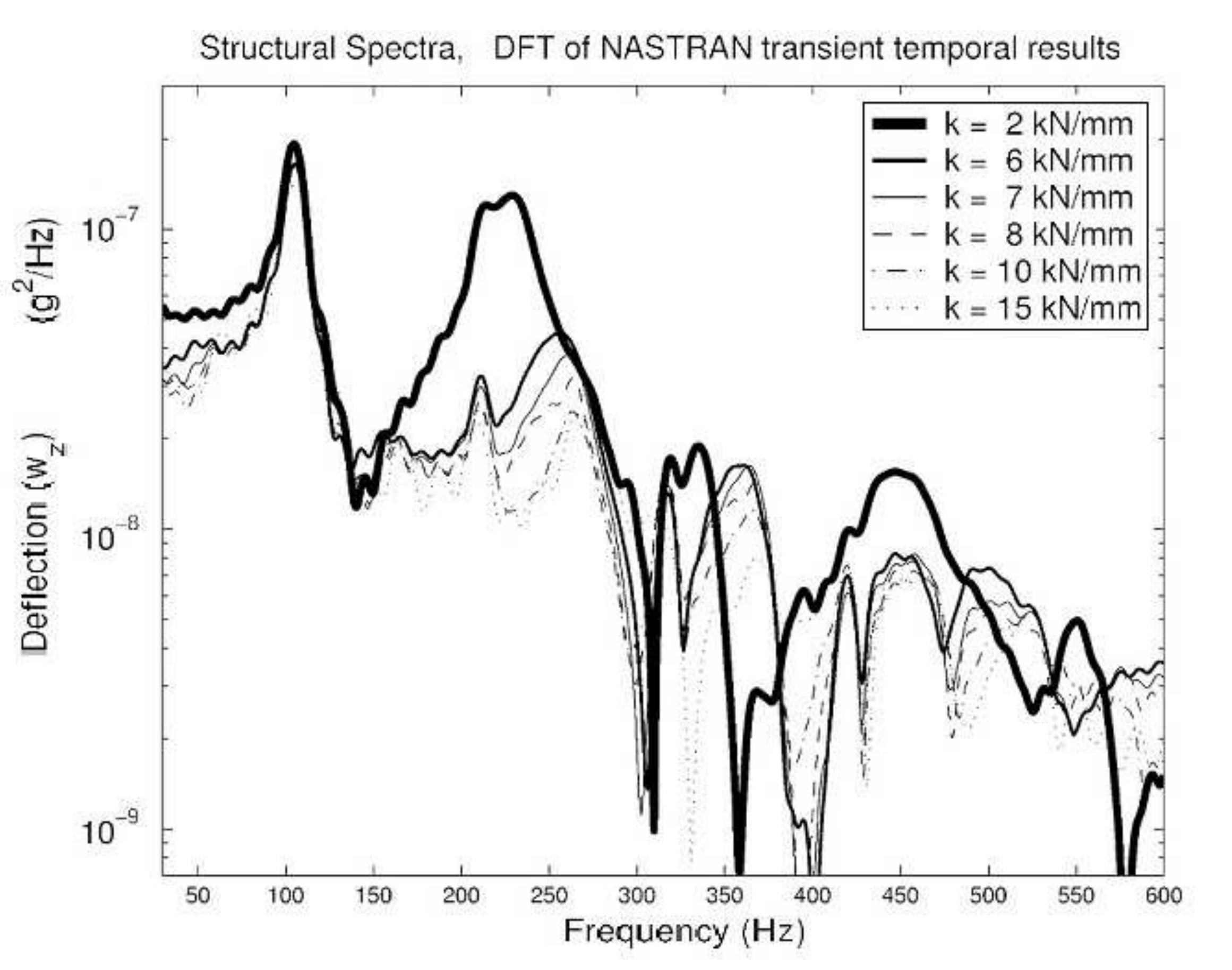}}
\caption[structuralSpectrum]{\label{impactResponseFRF} Spectra for armor-hull clattering systems, except for the symmetrical modes.  Results use different armor-hull baffle stiffnesses (see legend).}
\end{figure}

Using the observations above, a conjecture can be stated for the system restoring force of simple harmonic motion, based on the gradient of potential energy, along with the FEA simulation included herein (Figure \ref{impactResponseFRF}), including industrial experience, and the test data from the childhood of most of us.  While system energy decays with frequency ``on average," a plot of the spectrum within its lowest half dozen modes might not show the 1/$f$ phenomenon, but that the full pattern does.  Sometimes the spectrum plot is `in the trees' and the ``forest" of the 1/$f$ phenomenon over the entire large bandwidth spectrum is not seen within the zoomed-in window of smaller width.  For example, consider the `in the trees' ranges of 150-250 Hz and 400-500 Hz for the much lower fundamental frequency system of clattering armor in the FEA results of Figure \ref{impactResponseFRF}.

This figure plots several spectra of the simulated structure, 1 $\times$ 2 meter clattering plate (homogeneous rolled armor, discussed later).  There is one curve for each level of damping (see legend) and they show a \emph{different} method of inter-modal energy transport.  The 1/$f$ pattern shows up in its structural vibration spectrum.  A collection of the deformed shapes for transient analysis in the CSC thesis \cite{MSthesis:kobold06} shows how the transient result at each time step is a superposition of normal modes (eigenvectors) populated with energy according to the MPFs $\Phi(f)$. Those normal modes plots show how one mode at one frequency for one plate excites another mode at a different frequency on the other plate because of deflections that line up along the surface to contact the other plate at anti-nodes of a different mode. Again, the lower modes tend to be receiving modes.

% was: % \includegraphics{structuralSpectra7apr5compressed.eps}}

%\begin{figure}
%\resizebox{8.8cm}{!}{

% structuralSpectra2

%\caption[structuralSpectrum]{\label{impactResponseFRF} The spectrum of several armor-hull clattering systems except for the symmetrical modes.  The results for several with different armor-hull baffle stiffnesses (see legend).}

%%\caption[structuralSpectrum]{\label{impactResponseFRF} The spectrum of several armor-hull clattering systems drops off on the order of 1/$f$ except for the symmetrical modes.  Several different models are shown together with different armor-hull baffle stiffnesses (see legend) yet the changed configurations still have more energy in the lower frequency modes starting with the 108 hertz fundamental mode.}

Using appropriate structural damping, the frequency response function in Figure \ref{impactResponseFRF} is the result of an impulse load that rings all the modes of the FE model.  It appears to be natural that MPFs decrease with the mode number of the mode they characterize; in a time-averaged sense, the vector is an array of monotonically decreasing participation values usually measured as the strain-energy for each mode.  This is part of the reason for the ubiquitous 1/$f$ phenomenon.  At least for structural vibration \cite{book:FEAbyMacNeal94}, we can thank the Zienkiewicz section quoted above \cite[340, 341]{book:Zienkiewicz91} provides an explanation of at least some of this 1/$f$ fall-off.

\section{Simple harmonic motion, not so simple}

\label{SHO}

A spring-damper-mass \emph{single} DOF system (SDOF) provides a one dimensional model of the clattering armor plate which can provide analytical solutions to the vibration DEs.  Assume `small deflection,' where strain-energy density is low enough for linear elasticity assumptions (lack of permanent set in the structure being modeled).  Also assume that modes are nearly monochromatic functions of sines and cosines.

In a more formal development of the transfer of energy from one mode to another, Lord Rayleigh points out that pure sine or cosine waves do not exist in reality.  He derived the differential equations for Newton's theory for isothermal compressive-vacuum vibration\footnote{``Since the relation between the pressure [$p$] and the density [$\rho$] of actual gases is not that expressed in [ $ p=const-(u_o^2\rho_o^2 / \rho) $], we conclude that a self-maintaining stationary aerial wave is an impossibility, whatever may be the velocity $ u_o $ of the general current, or in other words that a wave cannot be propagated relatively to the undisturbed parts of the gas without undergoing an alteration of type.  Nevertheless when the changes of density concerned are small, [ $ p=const-(u_o^2\rho_o^2 / \rho) $] may be satisfied approximately; and we can see from [ $dp/d\rho=(u_o^2\rho_o^2 / \rho^2) $ ] that the velocity of stream necessary to keep the wave stationary is given by [ $ u_o=\sqrt{dp/d\rho} $ ] which is the same as the velocity of the wave estimated relatively to the fluid." \cite{book:RayleighTOS2}} resulting in this second order form \cite{book:RayleighTOS2}.  The footnote describes these generic variables in Equation \ref{eq:nonlinearWavEQ} for a wave in $y$ with respect to $x$ and time $t$.  These concepts are covered in engineering vibration textbooks \cite[p. 30]{book:Thomson88}.

\begin{equation} \label{eq:nonlinearWavEQ}
\bigg(\frac{dy}{dx}\bigg)^2\frac{d^2y}{dt^2} =
\frac{dp}{d\rho}\frac{d^2y}{dx^2}
\end{equation}

Lord Rayleigh then comments on the ability of simple harmonic motion to maintain shape in nature. The extent to which the mode shapes are not harmonic introduces a possible corruption of the pure energy-per-mode concept assumed by MPFs.  In the scientific literature a `mode' is derived from the parameters used for the study of statistics which include the median, mean, and the mode.  In a field where Gaussian distributions abound thanks to the physics described by the Central Limit Theorem,\footnote{The extent to which the literature supports the Central Limit Theorem can be found in papers from 1937 through 2003 with Uspensky \cite{book:Uspensky}, Landon and Norton \cite{inproc:Landon}, Khinchin \cite{book:Khinchin}, North \cite{TR:North}, and Le Cam \cite{ppr:LeCam}.} the mode is merely the location on the abscissa, in terms of the frequency for a spectrum, at the maximum of the response.  In this sense the typical emphasis on energy-in-mode used by modal engineers is appropriate \cite{ppr:AllemangMAC03} to identify the deformed \emph{shape} of the eigenvector for that approximate center frequency.  However, transfer functions are inappropriate for these nonlinear systems.  If transfer functions are modified as Bendat does for simple solitary nonlinear systems \cite{book:BendatNonlin98}, which is out of the scope of this work, there may be some consistency between the methods that would be useful to develop, separately.

\subsection{Contact nonlinear response}
\label{contactNonLin}

%%The author's CSC thesis \cite[App F]{kobold06} describes and plots the \emph{control law} as a sigmoid, using an arctangent function for the stiffness which is the slope of the load-deflection curve.  The control law is the bilinear contact stiffness model for a one gap element comprised of a large slope (very stiff) deflection curve for negative deflection which is compressive penetration, until lift-off where it becomes a small slope deflection curve, out to infinity (using a stabilizing small stiffness for numerical stability).  The CSC thesis describes the stability and application of this control law, and how the damping DOF and frequency DOF are no directly longer related.  The math follows Dr. Winthrop's December 2004 AFIT dissertation \cite[Eq 3.14]{Winthrop}. This nonlinear contact control law for the damped oscillator is only one of the suite of control laws Dr. Winthrop describes.

The author's CSC thesis \cite[App F]{MSthesis:kobold06} describes and plots the \emph{control law} as a sigmoid for the stiffness, which is the slope of the load-deflection curve.  The control law is a mathematical representation of the nonlinear stiffness associated with the contact state between armor and hull.  This classifier is an arctangent function, a smooth version of the typical bilinear gap element load-deflection table.  As the armor and hull make and break contact, the stiffness switches between high and low values of stiffness, respectively.  When the armor and hull are in contact, negative deflection (compressive penetration) results and there is a large stiffness due to the large slope of the deflection curve.  When the armor has separated from the hull, the nonlinear contact stiffness becomes a deflection curve of small slope.  This small value of extra stiffness provides computational stability.  The author's thesis \cite{MSthesis:kobold06} contains a description of the stability and application of the control law for this case, demonstrating that the damping DOF and frequency DOFs are no longer related.  The following analysis expands on one of Dr. Winthrop's control laws listed in his 2004 AFIT dissertation \cite{phdth:Winthrop04}.  His is a different control law, but the analysis methods are complementary.

%
%the bilinear contact stiffness model for a one gap element comprised of a large slope (very stiff) deflection curve for negative deflection which is compressive penetration, until lift-off where it becomes a small slope deflection curve, out to infinity (using a stabilizing small stiffness for numerical stability).  The CSC thesis describes the stability and application of this control law, and how the damping DOF and frequency DOF are no directly longer related.  The math follows Dr. Winthrop's December 2004 AFIT dissertation \cite[Eq 3.14]{Winthrop}. This nonlinear contact control law for the damped oscillator is only one of the suite of control laws Dr. Winthrop describes.

Phase space (state space) plots \cite[Fig 12]{MSthesis:kobold06} validate this model's behavior.  The arctangent function that does the switching in this simple closed form model appears in state space formulation as show below in Equation \ref{singleDOFrelation}, which has the constants listed as Equation \ref{oneDOFconstants1}.  The state space description for the single mass speed, $v$, and acceleration, $ \dot{v} $, shown in Equation \ref{singleDOFrelation} is a function of the open and closed stiffnesses, $ k_{\mbox{\scriptsize{open}}} $ and $ k_{\mbox{\mbox{\scriptsize{closed}}}} $, for a damping coefficient of $ \zeta = d/m $ in units of hertz, where $m$ is a point mass.

\begin{equation} \label{singleDOFrelation}
\begin{array}{l}
 \quad \dot{x} = v \\
  -\dot{v} = \frac{1}{2}\omega_o^2x(t)-\frac{1}{\pi}\tan^{-1}k_x x(t)
+ \frac{k_{\mbox{\scriptsize{open}}}}{k_{\mbox{\scriptsize{closed}}}} + \frac{d}{m}v \\
\end{array}
\end{equation}

This model of a ``welded-together" panel system\footnote{The term \emph{welded} assumes the plate deflections at both points are the same using an infinitely stiff connection. The closed gap stiffness used in this work is more like hard rubber \cite[App B.2.5]{MSthesis:kobold06}, but the effect is the same.} uses the parameters in Equation \ref{oneDOFconstants1}.  A ``manual," analytical calculation \cite[App C]{MSthesis:kobold06} helped validate the FEA results for the simulated vibration created before application of the laser vibrometry model.  Ngoya Pepela's optically measured modes provided qualitative support for the application of the laser vibrometry model that used the FEA results to modulate the probe beam.  This section and the Appendix describes the former, a structural SDOF model.

%\ref{oneDOFconstants1}.  The frequencies for different models where contact stiffness varied from model to model were in accord with three models: a manual calculation \cite[App C]{kobold06}, the FEA results for simulated vibration before application of the laser vibrometry model, and the optically measured modes.  All matched within engineering error.

\begin{equation} \label{oneDOFconstants1}
\omega_o = 2\pi 333.84, \qquad \frac{k_{\mbox{\scriptsize{open}}}}{k_{\mbox{\scriptsize{closed}}}} =
0.01, \qquad \frac{d}{m} = 0.02, \qquad k_x = 100  %
\end{equation}

In the arctangent gap model \cite[Fig 11]{MSthesis:kobold06} (not plotted here), the stiffness will transition at contact, $ x = 0 $.  Contact surfaces are imperfect due to microscopic protuberances that comprise the roughness of the surfaces.  As the two plates come into contact ($ x \leq 0 $) the roughness of the surfaces deform, compressing the protuberances, and a transition from low to high stiffness occurs rapidly in order to match the surrounding material. For a ``low" damping system, the damping is $\mu = 0.02$ kg/s and the open to closed dimensionless stiffness ratio is $k_{\mbox{\scriptsize{open}}}/k_{\mbox{\scriptsize{closed}}} = 0.01$.  The Appendix describes the dimensioned and dimensionless models.  A plot of speed versus gap opening in the thesis \cite[Fig 12]{MSthesis:kobold06} reveals that the high stiffness during compression of the base is a shallow orbit in $ x $ versus $ \dot{x} $ phase space. Changes in damping lead to changes in the range of the orbits during stabilization.  These control law formulations were implemented using a MATLAB system\cite{book:Polking99}.

%The transition to high stiffness at a gap opening of $ x \leq 0 $ in the arctangent gap model \cite[Fig 11]{kobold06} (not shown herein) is a smooth filet to model the collapse of surrounding protruding structure and imperfect contact plane. As \textit{x} decreases into negative territory, deforming the structural contact surface, the microscopic protuberances that comprise the surface roughness become compressed and the stiffness switches rapidly to match the surrounding material. For a ``low" damping system, the damping is $\mu = 0.02$ kg/s and the open to closed stiffness ratio is $k_{\mbox{open}}/k_{\mbox{closed}} = 0.01$.  Another plot in the thesis \cite[Fig 12]{kobold06} (also not shown) reveals that the high stiffness during compression of the base is a shallow orbit, a vertically oriented orbit in the $ x $ versus $ \dot{x} $ plot. Changes in (dimensionless) damping lead to changes in the range of the orbits during stabilization.  These control law formulations used a MATLAB system from Rice University \cite{Polking}.

Lyapunov function analysis shows the contact control law is asymptotically stable.  The Lyapunov function in this case is the total energy with simple damping loss.  In rare cases vibration energy gain to the hull upon which the armor plate clatters matches the cycle's damping energy losses and is in synch with plate contact. Persistent energy loss is due to structural damping, material heat losses due to bending, and fastener frictional losses.  The energy gain per cycle is less than the energy loss due to damping \cite[App F]{MSthesis:kobold06}. If the losses where small enough to just equal the gains, the system would be `stable in the sense of Lyapunov.'

%Lyapunov function analysis shows the contact control law is asymptotically stable.  The Lyapunov function in this case is the total energy with simple damping loss.  In rare cases vibration energy gain to the hull upon which the armor plate clatters matches the cycle's damping energy losses and is in synch with plate contact. Persistent energy loss is due to structural damping, a general model for material heat losses due to bending, and some fastener frictional losses.  The thesis \cite[App F]{kobold06} explains how the energy gain per cycle is less than the energy loss due to damping. If the losses where small enough to just equal the gains, the system would be `stable in the sense of Lyapunov' (isL).

This analysis above summarizes a clattering system.  Details of the physics of boundary conditions, initial conditions, and operating states are found in the thesis \cite[App F]{MSthesis:kobold06}.

\subsection{Laser vibrometry return from the probe beam}
\label{probedImaging}

%%% @@@@@ get an interface from the description of the
%%%simulation to this actual measurement:

The noise floor did not rise in Figure \ref{missedMode} but rather, the surrounding spatial areas had phase differences within the large spot size that, due to spatial averaging, underwent destructive interference in a bandwidth that removed most of the 1460 hertz mode.\footnote{Due to the location of Chladni lines of nodes for the 1460 Hz mode, the return contained Doppler shifts from different phases of the vibration for that mode.  Pepela's laser vibrometer \cite{MSthesis:NgoyaPepelaLaserVib03} used these Doppler shifts from different locations on the object that destructively interfered.  Other laser vibrometers measure speed with two pulses or use other techniques.}  This interference is a combination of optical phase shifts along-range caused by reflection from deflection amplitudes that are related to the \textit{structural} wavelength along the bar for each mode.

The purpose of the vibrating bar measurement was to show spectral elimination, but it also shows the discrete nature of high quality (low damping) structural modes.  The FEA of this system \cite{MSthesis:kobold06} complemented this laboratory measurement \cite{MSthesis:NgoyaPepelaLaserVib03} by showing that, while spectral elimination can occur with structures that might be constructed in the lab (one dimensional modes shapes are the simplest super-symmetric structures), it is impractical to build commercial structures that are substantially super-symmetric.

%The purpose of the vibrating bar measurement was to show spectral reduction, but it also shows the discrete nature of high quality (low damping) structural modes.  The laboratory metal strap structure did not have clattering; it was a high frequency doubly-clamped bar for the purpose of investigating spectral elimination in laser vibrometry.  The FEA in the CSC thesis \cite{kobold06} complemented this laboratory measurement \cite{NgoyaPepelaLaserVib} by showing that the spectral elimination was limited to structures that might be constructed in the lab, not commercial structures.  This 1-D strip is the simplest super-symmetric structure to fabricate.  Others are impractical or too expensive to survive approval.  The lab system used a non-imaging detector, similar to the alternative of two observation methods suggested in the CSC thesis.

The main result of the CSC simulations was that academic models can produce reductions of vibration sensitivity that theoretically verify part of Ngoya Pepela's spectral elimination thesis \cite{MSthesis:NgoyaPepelaLaserVib03}, within meaningful assumptions.  The single and two DOF models described here show why this is the case, theoretically.  A separate 2014 article \cite{ppr:insensitive} describes the lack of spectral elimination (or spectral reduction) for vehicle and other types of structures that are manufactured economically.

The lab vibration measurement shared some spectral reduction features seen in the nonlinear cross-spectral covariance simulation \cite{MSthesis:kobold06}.  The transient vibration results for the surface of the 3-D armor plate were the input for a MATLAB system that forms an image similar to one that a laser vibrometer would detect.  Commercial laser vibrometers typically only provide images processed with their proprietary system.  For the purpose of simulating the image of the vibrating plate, the optical model for this work used conventional Fresnel diffraction in a MATLAB system of functions.  The Fresnel propagation method introduced by Goodman \cite{book:GoodmanFO68} provides a computationally adequate method to propagate an image of the return from the target back to the detector.  In this sense, Pepela's laboratory measurement provided a validation for the FEA and vice versa.

%The lab vibration measurement shared some spectral reduction features seen in the nonlinear cross-spectral covariance simulation.  In the work for this CSC thesis \cite{kobold06} the transient vibration results on the surface of the 3-D armor plate were the input for a MATLAB system that forms an image related to those that a laser vibrometer would detect.  Commercial laser vibrometers do not usually provide the raw imagery to the user.  They provide some proprietary processed images.  For the purpose of simulating the image of the vibrating plate the CSC thesis used conventional Fresnel diffraction integrals.   The Fresnel propagation method introduced by Goodman \cite{GoodmanFO68} provides a computationally adequate method to propagate an image of the return from the target back to the detector.  Pepela's laboratory measurement provided a validation for the FEA and vice versa.  A MATLAB system of functions propagate the return from surface vibration due to a probe beam back onto a simulated sensor screen.

The laboratory measurements in Figures \ref{allModes} and \ref{missedMode} use a clamped-clamped bar that has a high frequency fundamental mode (compared to vehicle fundamental modes that are below 50 Hz).  The spectrum has not dropped off with frequency by 3 kHz because this is still the ``lower frequency region" for the structure, as discussed in the section \ref{ZienkiewiczEnergyFlow}.  It has a stiff fixity meant to test the laser vibrometer for structural wavelengths that would undergo a difference in phase across the beam spot size.

%The laboratory measurements in Figure \ref{allModes} and \ref{missedMode} use a clamped-clamped bar that has a high frequency fundamental mode (compared to vehicle fundamental modes that are below 50 Hz).  The spectrum has not dropped off with frequency by 3 kHz because this is still the ``lower frequency region" for the structure, as discussed in the section \ref{ZienkiewiczEnergyFlow}.  It has a stiff fixity meant to test the laser vibrometer for structural wavelengths that would undergo a difference in phase across the beam spot size.  Like most frequency response functions of common structures, if this spectrum were extended out past 10 kHz the 1/\textit{f} response fall-off would become apparent as seen in Figure \ref{impactResponseFRF} for the CSC simulation \cite{kobold06}.

The spectral elimination effect, seen mostly in(Figure \ref{missedMode}), provided laser vibrometer manufacturers information on what spot size to suggest or program for their probe beams.  They want to avoid such `spectral elimination.'  The simulation and analysis in the CSC thesis \cite{MSthesis:kobold06} shows users and manufacturers that, except for 1-D structures (bars), even specially manufactured structures are difficult to create in the super-symmetric form, and thus do not exhibit spectral elimination.

%The spectral elimination effect (Figure \ref{missedMode}) that was the subject of the 2003 bar vibration thesis \cite{NgoyaPepelaLaserVib} helps laser vibrometer manufacturers in the event users try to use their internal signal processing on an image with a \textit{large} spot probe beam.  Laser vibrometry manufacturer's may want to prohibit such `spectral elimination.'  The simulation done in the CSC thesis \cite{kobold06} does not have this missing mode problem by assuming manufactured targets are not super-symmetric.

A detailed study is available that has comparisons of both structural and optically sensed CSCs, to imaging versus non-imaging returns \cite{MSthesis:kobold06}.  The latter choice of observation type, non-imaging, analyzes a scalar signal composed of the spatial average of the entire image per time step.  The simulation analyzes the surface of homogeneous rolled armor (HRA) using only one scalar metric, radiant flux that is the spatial integration of the irradiance reflected from the HRA.  These coupled FEA-optics results, using non-imaging return, identify the modes on the target (HRA clattering on a hull), nearly as well as the imaging version of the CSC.

\subsection{Finite element modeling and analysis}

\begin{figure}
\resizebox{9cm}{!}{
\includegraphics{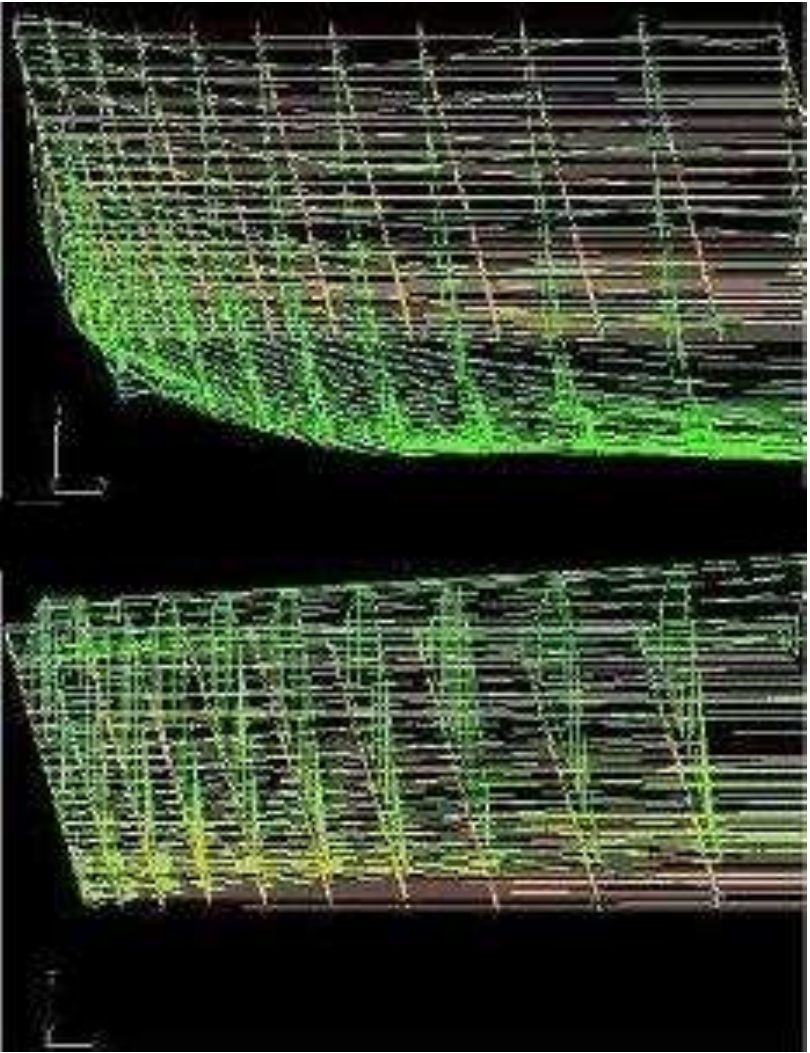}}
\caption[deformedShapeFEA]{\label{defShape} FEA of the deformed shape (green) for the clattering armor-on-hull model.  Exaggerated micron deflections would not be visible compared to the large physical geometry of the plates.  There is a 14.24 ms time difference (many time steps) between these two output time frames.  The views are \textit{into} the fore-shortened long direction (1 m) of the 0.5 m wide plate. The un-deformed shape is a light tan color.}
\end{figure}

The FEA produces a surface of vibration that is curved in 3-D.  The vibrational deflections are the input for the scripts and functions run within MATLAB that produce an optical simulation of the target's image.  Unlike the small and large spot sizes (Figures \ref{allModes} and \ref{missedMode}) in Ngoya Pepela's measurement \cite{MSthesis:NgoyaPepelaLaserVib03}, the remote sensing of the armor-plate clattering uses a complete coverage large spot size such that all points on the armor modulate the probe beam.  Assume the surrounding clutter to be time-gated or otherwise removed.  The input that the MATLAB script uses is the set of displacements to simulate images of the transient dynamical response of the surface of the clattering armor for hundreds of time steps in a duration of a few seconds.  One such `vibration deformed shape' is seen in Figure \ref{defShape}.

%\begin{figure}
%\resizebox{8.4cm}{!}{

% detect1740musecDefOnly

\begin{figure}
\resizebox{9cm}{!}{
\includegraphics{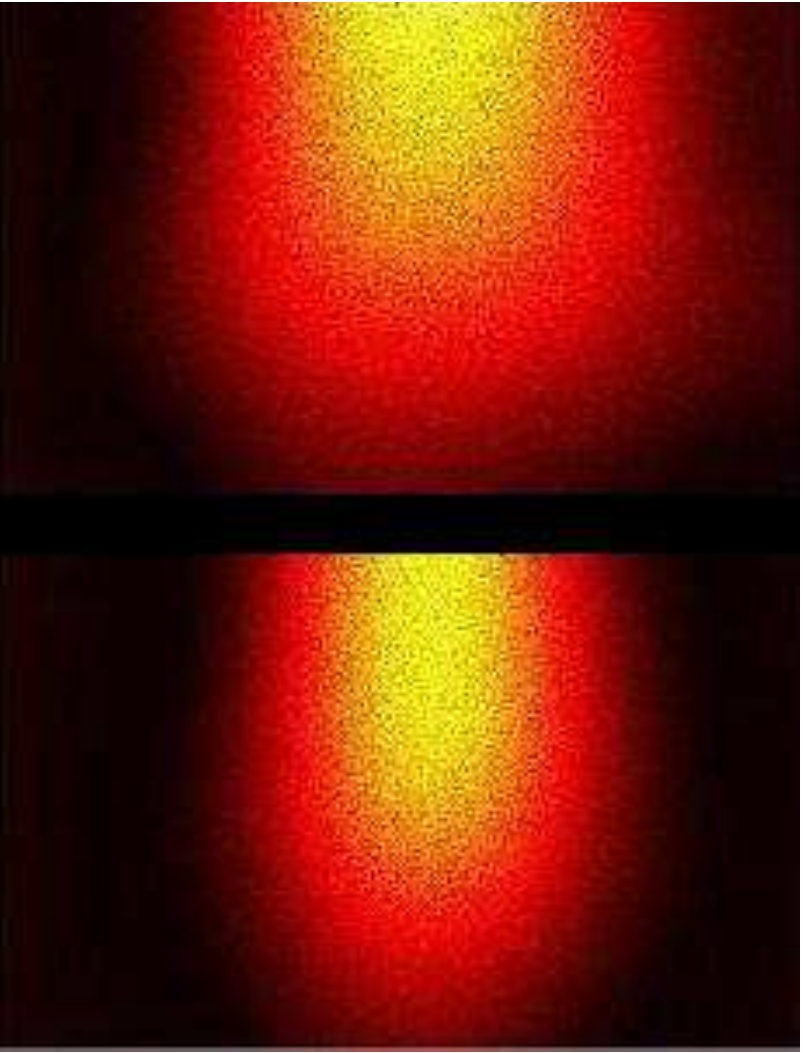}}
\caption[opticalImageHRA]{\label{imageOfReturn} Simulated magnitude of radiant flux image of the return after modulation by the vibrating plate in Figure \ref{defShape}.  The image horizontal width appears rotated 90 degrees from the long fore-shortened axis in the prior image in order to better display the 20 $\times$ 40 mm detector grid shape.  The maximum deflection ``hill" along the 1 m length shows up as a vertical maximum magnitude in the lower (later) pane of both figures.}
\end{figure}

The FEA representation of the vibration shape was applied to the MATLAB model of the probe beam as a phase modulation.  Using Fresnel diffraction \cite{book:GoodmanFO68}, the return was imaged onto the detector 4 km away.\footnote{For the computers at that time, even when running MATLAB simultaneously on several machines on the Air Force Institute of Technology cluster, there was a delicate balance between adequate structural grid densities and optical grid densities for the spatial Fourier transform that performs the Fresnel propagation of the return.  (It is more of a numerical modeling issue than a matter of compute power.)}  A 10 $\mu$m  probe wavelength satisfied the numerical requirements for results shown in Figure \ref{imageOfReturn}.
%
%\begin{figure}
%\resizebox{8cm}{!}{

% detect1740musecOptOnly

A spectrum of the time sequence of sensed images, such as the two shown in Figure \ref{imageOfReturn}, appears in Figure \ref{impactResponseFRF}.  More detailed plots are available \cite{MSthesis:kobold06}.

\section{Nonlinear eigen-states}
\label{nonLinEig}

Creating a simple model for a clattering structure may seem easy, until the fact that the eigenvalues must be nonlinear complex functions becomes apparent. A two degree of freedom (DoF) model developed in the Appendix herein was meant to validate the FEA.  This FEA validates a major dynamical 2 DOF analysis result: Antisymmetrical \emph{and} unsymmetrical modes increase in frequency for increases in contact stiffness.\footnote{The thesis\cite{MSthesis:kobold06} defines several forms of symmetry.  These symmetries include structural, clattering (as opposed to synchronized hull and armor), and modes similar to the deformed shape of the components to inertial loads.  The latter mode is similar to a mode shape (eigenvector) where the hull and armor move synchronized together, $w_{\mbox{\scriptsize{hull}}}(x, y) = w_{\mbox{\tiny{MRA}}}(x, y)$ from which comes the 1 degree of freedom (DOF) and 2 DOF models where the plates are averaged over $x$ and $y$ into two point masses.  The 1 DOF model grounds the ``hull" mass point to zero displacement, so it is removed in the 1 DOF model.} Investigation of these closed form simple damped sprung mass systems\footnote{`Sprung mass' systems are an automotive term for a systems where vibration is being isolated or suppressed.  At an academic or `free body diagram' level, they can sometimes be approximated by spring-mass-damper systems.} provides not only insight, but qualitative numerical validation. Whether an eigenvalue increases for anti-symmetric and un-symmetrical modes when gap-closed stiffness increases.  This simple model result provides another analytical basis for results seen in the FEA. Additionally, the behavior of the modes indicates that \emph{symmetrical} modes are better target identification features than un-symmetrical modes. Hence, there is a need to analyze the energy balance and stability \cite[App F]{MSthesis:kobold06} to validate the simple one and two DOF models as is summarized in the Appendix herein, including the time variation of amplitude and phase.

% was %   \includegraphics{realEigen2dof.eps}

%\begin{figure}
%\resizebox{0.52\textwidth}{!}{

% realEigen2dof2

\begin{figure}
\resizebox{0.8\textwidth}{!}{%
  \includegraphics{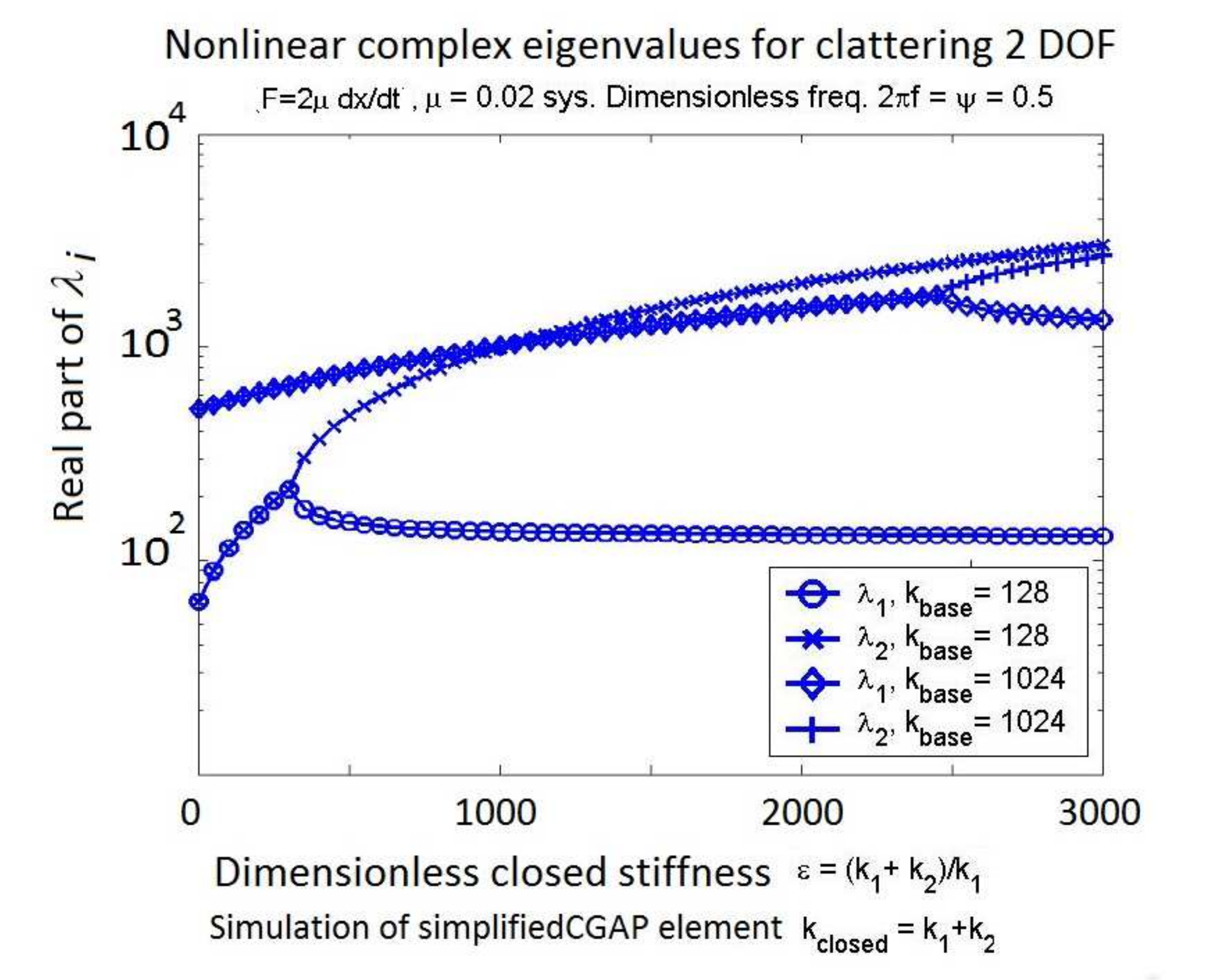}
}
\caption{\label{realEigen2dof} Real parts of the eigenvalues for the dimensionless DE show why higher-energy antisymmetric modes are less likely to be excited.  Above a $k_{\mbox{crit}}$ the symmetric (lower frequency branch) and antisymmetric (higher frequency branch) modes \cite[167]{book:Thomson88} for this two DOF problem break out into modes of well separated energy.}
\label{fig:1t-old}       % Give a unique label
\end{figure}

Using the SDOF model as a basis, the Appendix uses Equation \ref{controLawEq} to provide the 2DOF solution.  Equation \ref{controLawEq} may appear deceptively simple because the time variation of the amplitude and phase is not apparent until explicitly formed.  Eigenvalues $\lambda_i$ are complicated functions of natural frequencies of individual modes which require `mode tracking' \cite{ppr:modeTracking95} with respect to stiffnesses \cite{phdth:modeTracking93}, due to their transient nature in reality  \cite{ppr:modeTrackingAnderson84}.  Figure \ref{realEigen2dof} plots the real part of the dimensionless eigenvalues of Equation \ref{controLawEq} as a function of dimensionless frequency versus dimensionless closed stiffness $k_{\mbox{\scriptsize{closed}}}$, scaled by the hull stiffness $k_1$ for a closed stiffness $\epsilon^{\prime} \equiv 1/\epsilon = k_{\mbox{\scriptsize{open}}} /k_{\mbox{\scriptsize{closed}}}$ \cite{MSthesis:kobold06}.  Two models appear in Figure \ref{realEigen2dof}.  The hull system is eight times stiffer than the ``armor" system.

%
%Using the SDOF model as a basis, the analysis leading to Figure \ref{realEigen2dof} follows in the Appendix.  Figure \ref{realEigen2dof} plots real part of the eigenvalues of Equation \ref{controLawEq} as a dimensionless function of frequency (\cite{kobold06} summarized in the Appendix) versus dimensionless closed stiffness $k_{\mbox{closed}}$, scaled by the hull stiffness $k_1$ for a closed stiffness $\epsilon$.  Two models appear in the same plot.  One for a soft hull system, and another that is eight times stiffer.  Equation \ref{controLawEq} may appear deceptively simple because the time variation of the amplitude and phase is not apparent until explicitly formed in the Appendix.  Eigenvalues $\lambda_i$ are complicated functions of natural frequencies of individual modes which require `mode tracking' with respect to stiffnesses \cite{modeTracking93}, due to their transient nature in reality \cite{modeTrackingAnderson84}.

\begin{equation} \label{controLawEq}
m \frac{d^2x}{dt^2} + 2 \mu \frac{dx}{dt} + \omega_{sys}^2 \left( %
\frac{1}{2}+\frac{k_{\mbox{\scriptsize{open}}}}{k_{\mbox{\scriptsize{closed}}}}-\ %
\tan^{-1}\frac{100x}{\pi} \right) = 0 %
\end{equation}

From a chaos point of view, nonlinear attractors exist for the system defined by Equation \ref{controLawEq}.  We know this because the systems exist in reality and their vibrations modes do fluctuate, albeit not monotonically.  The nonlinear attractors are related to the underlying hull forcing functions (D'Alambert's forces due to vehicle inertial loads), the timing and shape of the clattering, the plate stiffnesses, the structural and added damping (e.g., washers or armor-hull batting), and the fixity of the joints.

Initially the two eigenvalues for the 2DOF system are equal.  As the closed stiffness increases, a critical value of stiffness, $k_{\mbox{crit}}$, is reached where the armor can no longer follow the hull.  It starts to clatter in an \emph{unsymmetrical} mode.  A 3-D contact surface cannot be anti-symmetric unless it is axisymmetric (1-D), as are the SDOF and 2DOF models.  When the stiffness exceeds $k_{\mbox{crit}}$ the two curves become distinct, branching into two separate paths.   This can be see for the plots of two models where the base stiffness is 128 and 1024 N/mm.  In the region before clattering, $\epsilon < k_{\mbox{crit}}$, the slope of eigenvalue curves, for the $8 \times k_{\mbox{\scriptsize{closed}}}$ higher base stiffness curve, decreases eight-fold compared to the smaller base stiffness.  At this higher 1024 N/mm base stiffness there is also an eight-fold increase in $k_{\mbox{crit}}$.  The intercept also increases 8-fold.  Derivations of these concepts and further results, including how the imaginary parts of the frequencies (not shown, Figure \ref{realEigen2dof} is the real part) contribute to the transfer of energy between the modes are developed in the Appendix.

%Individually, at some $k_{\mbox{crit}}$ the closed stiffness has increased to a level at which the outer plate (the armor) cannot follow the hull anymore.  So it starts to clatter.  That clattering is from the `unsymmetrical' mode.  (A 3-D contact surface cannot be anti-symmetric, but a 1-D models like the SDOF and 2DOF models can.)  Comparing the eigenvalue plots of the two models, the initial slope (dimensionless ``frequency" squared divided by dimensionless stiffness) decreases for the higher stiffness, $8 \times k_{\mbox{closed}}$.  At this higher stiffness there is also an eight-fold increase in $k_{\mbox{crit}}$.  The stiffness intercept, the fundamental `system' frequency squared, also increases 8-fold.  See the Appendix for derivations of these concepts, and further results, including how the imaginary parts of the frequencies (not shown in Figure \ref{realEigen2dof}) contribute to the transfer of energy between the modes.

\section{Analysis and conclusion}
\label{analysisConclusion}

A system model of the structural vibration of a contact-plate system was shown to have surface waves with low frequency modes that are effectively discrete modes of the PSD \cite{book:Thomson81}.  This clattering armor system provided an example of nonlinear response.  Both SDOF and 2DOF models derived in the Appendix have qualitatively verified the symmetrical versus unsymmetrical modes, and the higher frequency of the unsymmetrical clattering modes.  The full finite element model transient results also showed the tendency for vibration strain-energy to collect in the lower frequency modes, as most model engineers have seen \cite{inProc:Pininfarina}, and as was calculated analytically in the classical literature \cite[340, 341]{book:Zienkiewicz91}.

\begin{figure*}
\centerline{Table \ref{TableI}: Comparison of \textbf{Observational Characteristics} related to \emph{spectral elimination} (SE)}
\begin{tabular}{|r||l||l||l|}
\hline
\label{TableI}  Characteristic & 1. \textbf{Discrete modes} & 2. \textbf{Energy transitions} & 3. \textbf{Energy ordered by mode} \\
\hline
\hline
 Concept sketches & Figure \ref{discreteModeSketch} ($\Sigma_j^N E_j$) & Figure \ref{energyXfrSketch} ($ E_j \rightleftharpoons E_{j+1} $) & Fig. \ref{modalOrderSketch} ($ E_j \Longleftarrow E_{j+1} $) \\
\hline
 Dependency & High Q resonances & Clattering FEA model & Same deflection $\propto \omega^2$ at high \textit{f} requires $E_j \nearrow$ \\
\hline
 Spectral effect & Figure \ref{allModes} and \ref{missedMode} & Figure \ref{impactResponseFRF} & Damping, $\lim_{f\nearrow}c(f) \propto f$ eschews $E_{strain}$  \cite{book:Zienkiewicz91} \\
\hline
 Character & Low frequency & Nonlinear joints & Steady state vibration energy $E_i \propto 1/f$ \\
\hline
\hline
 Small spot size & Sees all modes & Blind to many $\Delta E_{jk}$ & No Spectral elimination (SE) \\
\hline
\hline
 Large probe spot & SE in bars & Sees more $\Delta E_{jk}$ & Some SE (spectral reduction) \\
\hline
 Super-symmetric & Some missed modes & Misses most $ E_j \rightleftharpoons E_{j+1} $ & Provides th. of SE \cite{ppr:insensitive}. Target rarely super-sym. \\
\hline
 Manufactured parts & Sees all modes, Fig \ref{impactResponseFRF} & FRF, coh(\textit{f}), MPFs & Typ. Figure \ref{impactResponseFRF} (super-sym. target is \emph{rare}) \\
\hline
\end{tabular}
\end{figure*}

%%\hline
%% & & & \\

(1) The lab tests on the doubly clamped bar show experimentally that high SNR modes are effectively discrete.  In the absence of substantial damping they are not just the maxima of a spectrum, but spikes in the spectral response \cite{MSthesis:NgoyaPepelaLaserVib03}.  The system in Figure \ref{allModes} has a signal-to-noise-ratio that is huge.  These massive SNRs show that the lower modes of structure, that are fairly high in frequency for this case due to the double clamped nature of the bar, can easily be considered discrete.  Hence, \emph{quod erat demonstrandum} (Q.E.D.), the observational characteristic (1) is demonstrated -- lower frequency modes are essentially discrete.

This analysis of nonlinear eigenvalues provides useful results for theory and simulation for common nonlinear structures.  Symmetrically moving parallel plates have more strain (and thus more strain-energy) than clattering plates where the unsymmetrical motion interrupts their nearly sinusoidal in time out-of-plane-motion to spew strain-energy into acoustics and even permanent set (deformation of the material).  Through simulation\footnote{Figure \ref{impactResponseFRF} shows that decreasing baffle stiffness between the plates (see its legend) drives unsymmetrical resonances lower.} and by experience the observational characteristic (1) appears true, that lower modes are effectively discrete for high quality systems (low damping), and that, when there are pathways (nonlinear joints) that allow energy transfer, (2) and (3) are in effect; damping and clattering help the energy flow into the lower modes from higher frequency modes -- although overall energy is lower as damping increases (see Figure \ref{impactResponseFRF}).  Most vehicle components have bolted or riveted joints that allow such energy transfer \cite{book:BendatNonlin98} as can be seen with plots of coherence spectra, coh(\textit{f}) \cite{book:BendatPiersol93}.

(2) The tools in industry that provide MPFs include mature engineering methods \cite{book:Thomson81} from the 1980's such as the FEA tools put in place, for example, by MacNeal-Schwendler Corporation engineers in NASTRAN \cite{inProc:MPFbyTedRose98}.  These MPFs stored in the FEA vector $\Phi$ can measure the energy transmitted between states (between eigenvectors of the system) where the state (mode) $i$ changes it energy with mode j. $\Delta\Phi_t(f_i) = \Phi_t(f_i) - \Phi_{t-1}(f_i) = -\Delta\Phi_t(f_j)$.  For surface waves on the ocean \cite{book:TalleyPickard11} this observation would include wave energy of large ocean swells driving up capillary waves and eventually creating foam as seen in large storms.  Energy transfers from one type of wave to another, Rossby-Kelvin, gravity-capillary, internal-surface, ..., and direction variations thereof, may prove useful.

Modal engineering for industry, especially for vehicles (i.e., ride-and-handling) is a trade secret endeavor.  Vibration spectral plots for vehicles found in the literature, such as the Pininfarina paper \cite{inProc:Pininfarina} are rare.  It makes use of analysis tools in order to calculate modal participation factors \cite{inProc:MPFbyTedRose98}.  These MPFs show the `participation' (energy per mode \cite{book:Thomson81}) using modes developed from normal modes that are FEA-produced eigenvectors.  The participation flows from energetic modes to lower energy modes similar to how large swells on the ocean in a storm are accompanied by unsettled surfaces, rather than smooth large waves that a tsunami has before it nears the shore \cite{book:TalleyPickard11}.  In the latter case, not enough time has passed to transfer the energy to other modes until the wave crashes on shore where sufficient coupling to other modes exists due to the constraints of the shore structures.

Therefore, due to (3) the energy ordering of modal states for complicated system synthesis models and the ordinary vehicles they represent, the work of vehicle design for these issues focuses on (2) transition probabilities\footnote{Transition probabilities are related to CSCs \cite{MSthesis:kobold06} and the coherence spectrum \cite{book:BendatPiersol93}.} and (1) energy levels $\Phi(f_i)$.  There are usually other types of oscillator interactions that produce an ordering of energy levels similar to that described by Zienkiewicz \cite[340-341]{book:Zienkiewicz91}.  These concepts are summarized in Table \ref{TableI}.

Pencil-thin probe beams like that used in the lab measurement have observational characteristics quite different from the large spot size used for the FEA-MATLAB model of fully illuminated clattering armor.  The former can be less susceptible to spectral elimination.  However, Mr. McKinley's observations discussed in section \ref{sect:measuredVibClamped$^2$} on page \pageref{sect:measuredVibClamped$^2$} show different modes appear and disappear for either change in spot size.  The latter low fidelity beam method is adequate for spectral identification of economically manufactured vehicles.  Such full coverage probe beams are less likely to be subject to spatial coherence issues or illuminate solely a node of the Chladni zone \cite{book:ChladniWiki}.  Pencil-thin beam returns fail to convey vibration modulation in this manner \cite{book:RayleighTOS1}.  For the large (full coverage) spot size, some of the beam will nearly always get through.

While these results are generic for plate structures, application to waves on the surface of a volume appear to fulfill similar behavior of (1) discrete modes, (2) modal participation dissipation, and (3) higher energy in lower frequency modes.  A tsunami has a much lower spatial frequency and temporal frequency than typical 2-4 meter gravity waves.  However, the larger Rossby and Kelvin waves are also candidates for the study of the extreme low frequency application of the principles observed in this article.

% the principles observed in this article in the author's dissertation proposal to FAU and NSWC PCD.  That dissertation seeks to characterize acoustic communications channels using the structure of the ocean surface waves above the acoustic channel.

\section{Appendix -- One and Two DOF contact eigenvalues}
\label{nonlinearEigenvaluesAppendix}

This Appendix summarizes the nonlinear dynamics of a simplified one-dimensional (1-D) and 2-D forms of a structural contact system,  its relationship to the full 3-D FEA, and in the end, test measurements.  The 2-D form provides for a difference in foundation stiffness, and transition to 3-D.

\subsection{Closed-form SDOF nonlinear contact response}
\label{closedFormSDOF}

This subsection displays the Mathcad\texttrademark output for the closed form solution to the damped SDOF oscillator meant to represent a lumped mass model of the HRA-hull contact vibration system. The mathematical derivations are based on the nonlinear solutions \cite{phdth:Winthrop04} with a simplification of the control law, $u(x)$, that models simple contact along the $x$-axis.

%\cite{phdth:Winthrop04} with a simplification of the control law (shown in the next subsection) that models simple contact. $x(t)$ of Equation \ref{x1DOFsolution} is Dr. Winthrop's relation \cite[Eq 3.14]{phdth:Winthrop04}.

%\subsection{Single DOF (SDOF) calculations}

The SDOF dimensionless system describes the effect of nonlinear contact stiffness from a solution composed of symmetric and antisymmetric one dimensional modes.  This SDOF solution applies to the symmetrical and unsymmetrical modes of the 3-D FE model for the HRA-hull system, respectively.  In 3-D the unsymmetrical modes are non-uniform, occurring when some parts of the armor is moving opposite to the hull (i.e. clattering).  For clarity 1-D modes are distinguished as symmetric or antisymmetric, while 3-D systems are labeled symmetrical\footnote{The symmetrical 3-D mode shows up in frequency response curves \cite{MSthesis:kobold06} but it is not perfectly synchronous across the surface, thus it is not a perfectly symmetric mode like the 1-D model.  It exists in 3-D the same way quantum mechanical modes exits - because of their energies - spikes in the spectra at the correct frequency, not because we saw them optically.  Although with FEA plenty of studies on modal decomposition are in the literature to reinforce the combination of modes that modal engineers use to analyze vibration.  The symmetric (1-D) and symmetrical (3-D) modes do not clatter while the other modes do.} and unsymmetrical.\footnote[2]{}  Application of derivatives to the dimensionless system is first made under the assumption that all variables are nonlinear functions of time.  Then after starting with a restricted case, the variables are brought into explicit nonlinear use, one at a time, to refine the calculation.

% \footnote[\ref{symmNote}]{}  % Sometimes this works, sometimes it does not.
% = 2  I just want to make a copy of the call-out to footnote \ref{symmNote}

%
%The SDOF calculations shown below, with suitable simplifications and assumptions, are used in the next subsection for the 2DOF solution. This SDOF description describes the effect of contact stiffness (bi-linear stiffness nonlinearity) on a solution composed of symmetric and antisymmetric one dimensional modes.  This solution applies to the symmetrical and unsymmetrical modes (respectively) resulting from the FE model for the tank hull and armor.  The SDOF model is a severely lumped-mass model. Different symmetry adjectives are necessary for the three dimensional FE model because clattering modes where the armor is moving opposite to the hull does not have the laterally spatial uniformity implied by the word `antisymmetric.'  Hence, for 3-D systems the latter symmetry adjectives are appropriate whereas for the 1-D system the former simpler words apply.

%At first the derivatives act on the dimensionless system assuming all variables are nonlinear.  After starting with a restricted case, we can bring the variables into explicit nonlinear use, one at a time, to refine the calculation.

The state-space representation \cite{book:ChenSlinSys} is shown here in its phase space form (location $x(t)$ and derivatives).\footnote{The state-space definition of the system combines the state $x(t)$ with its output $y(t)$ and changes in state ($\partial x/\partial t$).  These matrix relationships \cite{book:Brogan} of the state change vector ($\dot{x}$) and system output and its state and input ($u(t)$) are one of many sets of $A, B, C$, and $D$ matrices, easily confused with other field's ABCD systems such as optical ABCD ray matrices for laser resonator orientation \cite{book:laserSilfast,book:lasersSvelto98}.
\begin{list}{}
  \item $\dot{x} = Ax + Bu$  using the state and input matrices
  \item $y = Cx + Du$ via output and feed-through matrices
\end{list}
All matrices and variables can be functions of time.  Some definitions distinguish state space as discrete compared to continuous phase space. However the field of linear systems often uses continuous output and even state variables and the resulting Kalman filter is nearly ubiquitous \cite{book:ChenSlinSys}.}  With the definition $ \quad \cos \phi_{\psi\beta} \equiv \cos \left[ (\dot{\psi} t + \psi_o) t + (\dot{\beta} t + \beta_o) \right] $,\footnote{$\psi$ and $\beta$ are canonical rate terms for frequency and phase.  Their subscripted `o' terms are the initial frequency and phase values.  The `naught' terms have units of hertz and radians while the un-subscripted terms are time derivatives of frequency and phase.}

\begin{equation} \label{x1DOFsolution}
x(t) = (\dot{a}(t) t + a_o(t)) e^{-(\dot{\mu}(t) t + \mu_o(t))t}  %
\cos \phi_{\psi\beta}
\end{equation}

%$x(t)$ of Equation \ref{x1DOFsolution} is Dr. Winthrop's relation \cite[Eq 3.14]{phdth:Winthrop04}.
%
%Even the first derivative, Equation \ref{x1DOFsolution2}, is non-trivial.

%\begin{equation} \label{x1DOFsolution2}
%\dot{x}(t) = \frac{d}{dt} \left[ %
%(\dot{a} t + a_o) e^{-(\dot{\mu}(t) t + \mu_o(t))t} %
%\cos \phi_{\psi\beta} \right]
%\end{equation}

%
%\cos  \phi_{\psi\beta} %
%\cos(\Psi\Beta(t))%

%$$
%\qquad \times \cos \left[ %
%(\dot{\psi}(t) t + \psi_o(t)) t + (\dot{\beta}(t) t + \beta_o(t)) \right] %
%\right] %
%$$
%
%
%Even the first derivative of Equation \ref{x1DOFsolution}, is non-trivial.  These relations will look simpler after some assumptions are worked out below.
% MathCad allows factoring in several ways.  Collecting on ($\dot{x}(t) e^{(\dot{\mu} t + \mu_o) t}$) on $ \cos \phi $ and then on $ \sin \phi $, Equation \ref{xdotDOFexpression1} provides for a somewhat compact expression of the speed.

Even the first derivative of Equation \ref{x1DOFsolution}, is non-trivial.  Mathcad allows factoring in several ways.  Collecting on $ \cos \phi $ and then on $ \sin \phi $, Equation \ref{xdotDOFexpression1} provides a compact expression for the undamped speed in Equation \ref{xdotDOFexpression1}.

%\mbox{The ``undamped" speed is} \qquad %

\begin{equation} \label{xdotDOFexpression1}
\begin{array}{l}
\dot{x}(t) e^{(\dot{\mu} t + \mu_o) t} = \\
  \left( %
2 \dot{a} \dot{\mu} t^2 + ( 2 a_o \dot{\mu} + \dot{a} \mu_o) t +
 a_o \mu_o -\dot{a} \right) %
 \\
\qquad \times \cos \left( %
\dot{\psi} t^2 + (\psi_o + \dot{\beta}) t + \beta_o) \right) %
 \\
  - \left( %
(2 \dot{a} \dot{\psi} t^2 + ( 2 a_o \dot{\psi} + \dot{a} \psi_o +
\dot{a} \dot{\beta}) t
+ a_o \psi_o - a_o \dot{\beta} \right) %
\\
\qquad \times \sin \left( %
\dot{\psi} t^2 + (\psi_o + \dot{\beta}) t + \beta_o \right) %
 \\
\end{array}
\end{equation}

With more assumptions restricting the nonlinearity of the solution where appropriate, the acceleration is also collected on $ \cos \phi $ and then $ \sin \phi $ in Equation \ref{xddotDOFsolution} for this simplified expression:

\begin{equation} \label{xddotDOFsolution}
\begin{array}{r}
  \ddot{x} = \left( %
 - \dot{a} \psi_o t + [\dot{a} \mu_o - a_o \psi_o^2] \right) %
\cos (\psi_o t + \beta_o)
\\
-(\dot{a} \mu_o \psi_o t + a_o \mu_o
\psi_o ) \sin (\psi_o t + \beta_o)
  \\
  \forall \qquad \dot{\mu} = 0, \quad  \dot{\beta} = 0, \quad \dot{\psi} = 0 \\
\end{array}
\end{equation}

The first order nonlinear solution in Equation \ref{xddotDOFsolution} uses a constant amplitude, \textit{a} = constant.  While not explicitly a function of time, \textit{a} has a constant time rate of change, $ \dot{a} $.  A further order of nonlinearity to allow the amplitude change rate to be a function of time would follow the above assumption with the use of $ \dot{\psi} $, the time rate of change of a dimensionless frequency.  The following subsections describe how Figure \ref{realEigen2dof} shows that while the stabilizing stiffness increases above the critical lift-off frequency, symmetric modes remain at a constant frequency while antisymmetric modes increase in frequency.  This 1-D behavior implies the same effect for 3-D modes, symmetrical and unsymmetrical, as is seen in the FEA results \cite{MSthesis:kobold06}.

\subsection{Closed-form Mathcad 2 DOF contact}
\label{closedFormMathCad2dof}

%The subsections to follow display Mathcad symbolic output for the two DOF (2DOF) model.  They are the closed form solutions for a damped 2DOF oscillator.  $k_x$ is the ``rate" that defines the sharpness of the contact.

Mathcad symbolic equations in the following calculations comprise the closed form solutions for a damped two DOF (2DOF) contact vibration model.  Stiffness $k_x$ is the ``rate" that defines the sharpness (hardness) of the contact.

%%This nonlinear 2DOF model represents a severely lumped mass model of the HRA-hull vibration. These dynamical equations are an extension of the SDOF relations developed in the prior subsection based on the nonlinear solutions in Dr. Major Winthrop's dissertation \cite{Winthrop} with selection of a ``control law" that models simple contact.  $k_x$ is the ``rate" that defines the sharpness of the contact.  $x(t)$ is Dr. Winthrop's dimensionless contact distance \cite[Eq 3.14]{Winthrop}.

\subsection{Two DOF DE's and solutions}

Equation \ref{DEfor2DOF} is a dimensioned form of the two DOF damped oscillator DE where $P$, $m$, $k$, \textit{d}, $\epsilon$, $\xi$, and \textit{t} are applied force, outboard mass, foundation stiffness, damping, control law stiffness, surface displacement, and time.

\begin{equation} \label{DEfor2DOF}
\begin{array}{l}
  \left(%
\begin{array}{c}
  P_1(t) \\
  0 \\
\end{array}%
\right) = \left(%
\begin{array}{cc}
  m_1 & 0 \\
  0 & m_2 \\
\end{array}%
\right) \frac{d^2}{dt^2} \left(%
\begin{array}{c}
  \xi_1 \\
  \xi_2 \\
\end{array}%
\right) + %
\\
 \left(%
\begin{array}{cc}
  d_1 & 0 \\
  0 & d_2 \\
\end{array}%
\right) \frac{d}{dt} \left(%
\begin{array}{c}
  \xi_1 \\
  \xi_2 \\
\end{array}%
\right) + \\
  \Bigg[ \left(%
\begin{array}{cc}
  1 & 1 \\
  1 & 1 \\
\end{array}%
\right) + \left(%
\begin{array}{cc}
  \epsilon_1 & 0 \\
  0 & \epsilon_2 \\
\end{array}%
\right) \bigg(\frac{1}{2} - \frac{1}{\pi} \left(%
\begin{array}{cc}
  \arctan(r \xi_1) & 0 \\
  0 & 0 \\
\end{array}%
\right) \bigg) \Bigg]%
\\
\times \left(%
\begin{array}{cc}
  k_1 & -k_1 \\
  -k_1 & k_1 + k_2 \\
\end{array}%
\right) \left(%
\begin{array}{c}
  \xi_1 \\
  \xi_2 \\
\end{array}%
\right) \\
\end{array}
\end{equation}

The dimensionless ratio $ k = (k_1+k_2) / k_1$ is the stiffness from the second oscillating point mass (the 1-D representation of the armor) to the base of the system as a whole, which includes the hull for this SDOF system. The hull is freed to oscillate in the 2DOF system.  The dimensioned variable $ \xi $ is the gap opening in the dimensionless \textit{x} direction.  The dimensioned control law for structural contact for this two DOF problem is the displacement $ u(\overrightarrow{\xi}) $ of Equation \ref{2DOFcntrLaw}.  The \emph{dimensionless} stiffness from the first oscillator to ground is unity.

%\textit{k} will be the dimensionless stiffness ratio for the second oscillator, $ k = (k_1+k_2) / k_1$ which is the stiffness of the base of the system.  $ \xi $ indicates the dimensioned form of gap opening

$ \xi $ indicates the dimensioned form of gap opening in the dimensionless \textit{x} direction.  The dimensioned control law for structural contact for this two DOF problem is the displacement $ u(\xi) $ of Equation \ref{2DOFcntrLaw}.

\begin{equation} \label{2DOFcntrLaw}
\mbox{Dimensioned} \qquad u(\overrightarrow{\xi}) = \Bigg[ \frac{1}{2} - \frac{1}{\pi} \left(%
\begin{array}{cc}
  \arctan(k_x \xi_1) & 0 \\
  0 & \arctan(k_x \xi_2) \\
\end{array}%
\right) \Bigg]
\end{equation}

The use of Winthrop's method \cite{phdth:Winthrop04} on Equation \ref{DEfor2DOF} produces the dimensionless DE in Equation \ref{dimlessDEfor2DOF}.  The dimensionless applied load is $F_{\mbox{applied}}$ and the dimensionless stiffness only applies to DOF one: $\overrightarrow{\epsilon} = [\epsilon_1; 0]$.

\begin{equation} \label{dimlessDEfor2DOF}
\begin{array}{r}
 \mbox{Dimension\textbf{less}} \qquad
\left(%
\begin{array}{l}
 F_{\mbox{applied}}(t) \\
  0 \\
\end{array}%
\right) = \qquad \qquad \qquad \qquad
\\
\\
\begin{array}{r}
\left(%
\begin{array}{cc}
  1 & 0 \\
  0 & 1 \\
\end{array}%
\right) \ddot{\overrightarrow{x}} + 2 \left(%
\begin{array}{cc}
  \mu_1 & 0 \\
  0 & \mu_2 \\
\end{array}%
\right) \dot{\overrightarrow{x}}%
 +  \left(%
\begin{array}{cc}
  1 + \epsilon_1 u(x_1) & -1 \\
  -1 & k \\
\end{array}%
\right) \overrightarrow{x}
\end{array}%
\end{array}%
\end{equation}

%%\end{array}% \\  \begin{array}{r}
%$
%\left(%
%\begin{array}{cc}
%  1 & 0 \\
%  0 & 1 \\
%\end{array}%
%\right) \ddot{\overrightarrow{x}} + 2 \left(%
%\begin{array}{cc}
%  \mu_1 & 0 \\
%  0 & \mu_2 \\
%\end{array}%
%\right) \dot{\overrightarrow{x}}%
% +  \left(%
%\begin{array}{cc}
%  1 + \epsilon_1 u(x_1) & -1 \\
%  -1 & k \\
%\end{array}%
%\right) \overrightarrow{x} %
%$

%%%$
%%%\left(%
%%%\begin{array}{cc}
%%%  1 & 0 \\
%%%  0 & 1 \\
%%%\end{array}%
%%%\right) \ddot{\overrightarrow{x}} + 2 \left(%
%%%\begin{array}{cc}
%%%  \mu_1 & 0 \\
%%%  0 & \mu_2 \\
%%%\end{array}%
%%%\right) \dot{\overrightarrow{x}}%
%%%\\
%%% +  \left(%
%%%\begin{array}{cc}
%%%  1 + \epsilon_1 u(x_1) & -1 \\
%%%  -1 & k \\
%%%\end{array}%
%%%\right) \overrightarrow{x} %
%%%$

Equation \ref{assumedSolutionX} uses the vector $ \overrightarrow{x} = [x_1 \quad x_2]^T $, the dimensionless location of the two masses where $ x_2 - x_1 $ is the gap opening.  Continuing with Winthrop's assumptions \cite{phdth:Winthrop04} assumes a straightforward solution. Equation \ref{assumedSolutionX} shows all the variables that vary with time.

%This relation uses the vector $ \overrightarrow{x} = [x_1 \quad x_2]^T $, the dimensionless location of the two masses where $ x_2 - x_1 $ is the gap opening.  Continuing with assumptions outlined in the Winthrop dissertation \cite{phdth:Winthrop04}, assume a straightforward solution. Equation \ref{assumedSolutionX} shows the remaining variables that vary with time.

%%Subsequent assumptions listed in Equation
%%\ref{LinearityAssumptions2DOF} will compare keeping some of these
%%variables constant to making the two DOF components equal.

\begin{equation} \label{assumedSolutionX}
x_i = a_i(t) e^{\mu_i(t) t} \cos(\psi_i(t) t + \beta_i(t))
\end{equation}
%
%Substituting assumed solutions from Equation \ref{assumedSolutionX} into Equation \ref{dimlessDEfor2DOF} results in a Special Eigenvalue Problem (SEVP).  The dimensionless system frequencies $ \psi_i $ are functions of the individual frequencies $ f_i = \omega_i /2\pi $ from $ \omega_i^2 = (k_i / m_i) - 2\zeta_i^2 $ where $ \zeta_i = d_i / 2m_i $.  (See the details in \cite{phdth:Winthrop04}.)  In order to assist in making the solutions

Substituting assumed solutions from Equation \ref{assumedSolutionX} into Equation \ref{dimlessDEfor2DOF} results in a Special Eigenvalue Problem (SEVP).  The dimensionless \emph{system} frequencies $ \psi_i $ are functions of the \emph{individual} frequencies $ f_i = \omega_i /2\pi $ from $ \omega_i^2 = (k_i / m_i) - 2\zeta_i^2 $ where $ \zeta_i = d_i / 2m_i $ from Equation \ref{DEfor2DOF}.  Two assumptions help make the solutions tractable.  First, the removal of the driving load provides the homogeneous solution.  Second, variables dependent on time and space (uniformity with respect to location) vary differently.  To first order the time variation of the dimensionless frequency, $ \psi $, is the derivative in this nonlinear system that has the largest effect on the solution \cite{phdth:Winthrop04}.  Equation \ref{LinearityAssumptions2DOF} summaries some of the simplifications:

\begin{equation} \label{LinearityAssumptions2DOF}
\begin{array}{r@{\quad}}
  a_1(t) = a_2(t) = constant = a \\
  \mu_1(t) = \mu_2(t) = constant = \mu \\
  \beta_1(t) = \beta_2(t) = constant = \beta \\
  F_{\mbox{applied}}(t) = 0 \\
\end{array}
\end{equation}

%%
%%\begin{equation} \label{LinearityAssumptions2DOFwords}
%%\begin{array}{l}
%%  \mbox{Constant amplitude `a' cancels out of the DE.} \\
%%  \mbox{Uniform damping $ \mu $ is for simplicity.} \\
%%  \mbox{But uniform phase $\beta$ is realistic.} \\
%%  \mbox{And finally, solve for free vibration first.} \\
%%\end{array}
%%\end{equation}

In words, these simplifications are:

\begin{itemize}
  \item Constant amplitude, \textit{a}, cancels out of the DE.
  \item Uniform damping, $ \mu $, is for simplicity.
  \item But uniform phase, $\beta$, is realistic.
  \item Solve for free vibration first (homogeneous solution)
\end{itemize}

Using a new variable for dimensionless phase, $ \phi = \psi t + \beta $, to simplify the state space gap opening (Equation \ref{x1DOFsolution} simplified in Equation \ref{assumedSolutionX}), the solution starts with a definition of the speed (Equation \ref{FirstOrder2DOFspeed}) and acceleration (Equation
\ref{FirstOrder2DOFaccelDef}), where coefficients are maintained as variables of time in the derivatives.

\begin{equation} \label{FirstOrder2DOFspeed}
\dot{x} = -\Big[ a\mu e^{-\mu t} \cos \phi + a\psi e^{-\mu t} \sin
\phi  \Big]
\end{equation}

Maintain the coefficients as variables of time in the subsequent derivative:

\begin{equation} \label{FirstOrder2DOFaccelDef}
\ddot{x} =  a (\mu^2 - \psi^2) e^{-\mu t} \cos \phi + a (\mu \psi
+ \mu \psi) e^{-\mu t} \sin \phi
\end{equation}

%Using the format of Equation \ref{dimlessDEfor2DOF}, the resulting dimensionless DE in Equation \ref{FirstOrder2DOFdimless2} below starts to take the form of an SEVP (Equation \ref{FirstOrder2DOFmatrixEq}), easily solved by eigenvalue methods.  We move representation of the time derivatives into the coefficients of the vectors in operator form so that the vector $\overrightarrow{x} = [x_1 \quad x_2]^T $ factors out.  Here we carry the decay (extinction) constant and frequency, $ \mu $ and $ \psi $ respectively, until getting the formal solution (``nonlinear" eigenvalues and eigenvectors).  A purely imaginary frequency $ \exp(if't) =  \exp(i[i\mu]t) $ is an exponential term in $ \exp(-\mu t) $.  In this case the sign of $ \mu $ indicates damping.  The ensuing 1-D math and descriptions provide insight into the dynamics and validate the transient `nonlinear' modes output by the 3-D FEA modes that exist in reality.

Using Equations \ref{FirstOrder2DOFspeed} and \ref{FirstOrder2DOFaccelDef}, time derivatives formed below come from application of these differential operators:

\begin{equation} \label{eq:diffOperatorsPiOmega}
\begin{array}{l}
\Pi_i \equiv (\mu_i^2 - \psi_i^2) + 2\mu_i\psi_i\arctan \phi_i \\
\\
\Omega_i \equiv 2\mu_i \bigg[-\mu_i - \psi_i\arctan \phi_i \bigg]
\end{array}
\end{equation}

%%Equations \ref{tDerivativePi} and \ref{tDerivativeOmega}:
%%
%%[I would use separate equations, but that makes almost half a page of white space because of how LaTeX formats these two small one-line expressions.]
%%
%%\begin{equation} \label{tDerivativePi}
%%\Pi_i \equiv (\mu_i^2 - \psi_i^2) + 2\mu_i\psi_i\arctan \phi_i
%%\end{equation}
%%
%%\begin{equation} \label{tDerivativeOmega}
%%\Omega_i \equiv 2\mu_i \bigg[-\mu_i - \psi_i\arctan \phi_i \bigg]
%%\end{equation}

From Equation \ref{dimlessDEfor2DOF} the resulting dimensionless DE in Equation \ref{FirstOrder2DOFdimless2} below starts to take the form of an SEVP (Equation \ref{FirstOrder2DOFmatrixEq}), easily solved by eigenvalue methods.  The time derivative operators $\Pi$ and $\Omega$ (defined above) allow the vector $\overrightarrow{x} = [x_1 \quad x_2]^T $ to factor out.  The damping constant, $ \mu $, and dimensionless frequency, $ \psi$, remain in the calculation until the ``nonlinear" eigenvalues and eigenvectors are developed in the formal SEVP solution.  A purely imaginary frequency $ \exp(if't) =  \exp(i[i\mu]t) $ is in the exponential term in $ \exp(-\mu t) $ where the sign of $ \mu $ indicates damping.  The ensuing 1-D analysis provides insight into the dynamics and validates the transient `nonlinear' modes output by the 3-D FEA, modes that exist in reality.

\begin{equation} \label{FirstOrder2DOFdimless2}
\begin{array}{l}
  \left(%
\begin{array}{cc}
 \Pi_1  & 0 \\
 0 & \Pi_2 \\
\end{array}%
\right) \left(%
\begin{array}{c}
  x_1 \\
  x_2 \\
\end{array}%
\right)
 + \\
  \left(%
\begin{array}{cc}
\Omega_1  & 0 \\
 0 & \Omega_2 \\
\end{array}%
\right) \left(%
\begin{array}{c}
  x_1 \\
  x_2 \\
\end{array}%
\right) + \\
  + \left(%
\begin{array}{cc}
  1 + \epsilon_1 u(x_1) & -1 \\
  -1 & k \\
\end{array}%
\right) \left(%
\begin{array}{c}
  x_1 \\
  x_2 \\
\end{array}%
\right) = \left(%
\begin{array}{c}
  0 \\
  0 \\
\end{array}%
\right) \\
\end{array}
\end{equation}

These relations use a different dimensionless stiffness, $ \epsilon \equiv \epsilon_1 = k_{\mbox{\scriptsize{closed}}} / k_{\mbox{\scriptsize{open}}} $ where $ k_{\mbox{\scriptsize{closed}}} = k_2$ and $ k_{\mbox{\scriptsize{open}}} \ll k_2$ for numerical stability. Equation \ref{FirstOrder2DOFdimless2} is of the SEVP format as shown below in
Equation \ref{FirstOrder2DOFmatrixEq}.

\begin{equation} \label{FirstOrder2DOFmatrixEq}
\left(%
\begin{array}{cc}
  A_{1,1} & A_{1,2} \\
  A_{2,1} & A_{2,2} \\
\end{array}%
\right) \left(%
\begin{array}{c}
  x_1 \\
  x_2 \\
\end{array}%
\right) = \left(%
\begin{array}{c}
  0 \\
  0 \\
\end{array}%
\right)
\end{equation}

Equation \ref{FirstOrder2DOFmatrixAdef} shows the format of the
system matrix A for submission to an eigen-solver.  Some of the
terms in $ A_{1,1} $ and $ A_{2,2} $ were kind enough to cancel.

\begin{equation} \label{FirstOrder2DOFmatrixAdef}
\begin{array}{l}
\mathbf{A} = \left(%
\begin{array}{cc}
  A_{1,1} & A_{1,2} \\
  A_{2,1} & A_{2,2} \\
\end{array}%
\right) = \\%
\\
\qquad \qquad
\left(%
\begin{array}{cc}
  -\mu_1^2 - \psi_1^2 + 1 + \epsilon u(x_1) & -1 \\
  -1 & -\mu_2^2 - \psi_2^2 + k \\
\end{array}%
\right)
\end{array}%
\end{equation}

%$
%\qquad
%\left(%
%\begin{array}{cc}
%  -\mu_1^2 - \psi_1^2 + 1 + \epsilon u(x_1) & -1 \\
%  -1 & -\mu_2^2 - \psi_2^2 + k \\
%\end{array}%
%\right)
%$

%
%Calculation of the determinant of \textit{A} with the 1997 version of Mathcad provides solutions to the characteristic polynomial of the system.\footnote{Mathcad '97 has a small Maple kernel that is far easier to use than recent 2016 versions - A 15 minute great solution versus days of dealing with customer support and still not quite getting the better solution.}  So there are two sets of solutions: free classical vibration, and the vibration solution where the parts are welded together (modelled as ``welded" is equivalent to epoxied with hard rubber in the FE model \cite{MSthesis:kobold06}).  The latter reduces to the former for slight contact, $u$ = 0, before compression ($u < 0$), as defined in Equation \ref{2DOFcntrLaw}.  Therefore, the ``welded" solution gives an indication of the dynamics by investigating the behavior of variations in control law $u$ from zero to one. The switching control law for the contact, $u(x)$, helps

%% Add a space after the equation with a null "equation"

$ $

Solution of the characteristic polynomial, $ |[A] - \lambda [I]| = 0 $, provides the eigenvalues of the system where both DOF have equal displacements.  Using a Mathcad `97 symbolic solver\footnote{The 1997 version of Mathcad has a small Maple kernel that is easier to use than recent versions.} two solutions are available:  (1) Free classical vibration, and (2) a "welded" system vibration solution.  The latter welded components (the two masses) is equivalent to being epoxied with hard rubber \cite{MSthesis:kobold06}.  As defined in Equation \ref{2DOFcntrLaw}, the welded model reduces to free vibration for slight contact, $u$ = 0, just before compression when \textit{u} turns negative.  The system dynamics are revealed via the \emph{welded} solution as the control law is varied in the range $0 \leq u \leq 1$. The switching control law, $u(x)$, determines which frequency is active, switching from one frequency value for the open gap state to another for closed gap state.  The nonlinear FEA results for 3-D show that both modes are active at the same time at different locations on the surface.  Spectral energy flows into and out of both `open gap' and `closed gap' modes when averaged over many cycles of time.  The contact gap system is a time composite system with dynamics that are more easily modeled using modal analysis techniques. The noise and vibration industry developed these techniques with nonlinearities such as contact in mind \cite{phdth:Allemang80,inbook:HarrisVib96,inbook:HarrisModal96,inbook:HarrisShock96}.

%The energy flow oscillation, as the system states evolve in an oscillatory manner, might be a useful model for other systems where the physics may appear to forbid energy level ordering.  Such disorder may happen on a time frame we are as yet unaware of.  However, for vehicle structures, experience indicates that the oscillation period is usually on the order of a minute to an hour.

%States evolve in an oscillatory manner.  This energy flow between alternately created and destroyed states might be a useful model for other systems where the physics appears to forbid continuous energy level ordering.  Such disorder may happen on a different time frame.  For vehicle structures, experience indicates that the oscillation period is usually on the order of a minute to an hour.

States evolve in an oscillatory manner.  This energy flow between alternately created and destroyed states might be a useful model for other systems where the physics appears to forbid continuous energy level ordering.  For vehicle structures, experience indicates that the oscillation period is usually less than an hour, on the order of minutes.

\subsection{Two DOF nonlinear contact eigenvalues}
\label{closedForm2dofNonLin}

%The eigenvalues in Equation \ref{FirstOrder2DOFeigenvalues} are therefore a dual set for zero and nonzero control law values, $u(x)$ for an open and closed gap, respectively.  In the dimensionless system, the non-contact state has unity stiffness in DOF 1 and stiffness of k for DOF 2.  But DOF 1 adds to the stiffness when in contact, so that base plus contact stiffness becomes $ 1 + \epsilon $.  This is the FEA maxim that `stiffnesses add" \cite{misc:aero51xAnderson93}.  The derivation of the dimensionless ``eigenvalues" $\lambda_i$ in Equation \ref{FirstOrder2DOFeigenvalues} came from the Mathematica results \cite{MSthesis:kobold06}.

Eigenvalues in Equation \ref{FirstOrder2DOFeigenvalues} are a dual set for the control law, $u(x)$, for an open and closed gaps.  In the dimensionless system \emph{for the open gap state} the stiffness for DOF 1 is normalized to unity and DOF 2 has a stiffness of $k = 1 + (k_2/k_1)$.  When the system goes into contact $u(x)$ adds $\epsilon$ to DOF 1 which becomes $ 1 + \epsilon $, as seen in Equation \ref{dimlessDEfor2DOF} and \ref{FirstOrder2DOFdimless2}.  This addition uses the FEA maxim that ``stiffnesses add" \cite{misc:aero51xAnderson93}.  The total closed stiffness is $ 1 + \epsilon + k$.  The derivation of the complete dimensionless ``eigenvalues" $\lambda_i$ in Equation \ref{FirstOrder2DOFeigenvalues} are Mathematica results for the solution of the eigenvalue problem described in Equations \ref{FirstOrder2DOFmatrixEq} and \ref{FirstOrder2DOFmatrixAdef} above \cite{MSthesis:kobold06}.

\begin{equation} \label{FirstOrder2DOFeigenvalues}
\begin{array}{r}
\left(%
\begin{array}{c}
  \lambda_1 \\
  \lambda_2 \\
\end{array}%
\right) = %
 \frac{1}{2} \left(%
\begin{array}{c}
  (1 + \epsilon + k)-2(\mu^2 + \psi^2) - \sqrt{\kappa} \\
  (1 + \epsilon + k)-2(\mu^2 + \psi^2) + \sqrt{\kappa} \\
\end{array}%
\right) \\
 \qquad
 \kappa \equiv 5 - 2 \epsilon + \epsilon^2 + 2 k - 2 \epsilon k - k^2 \quad (\epsilon \equiv \epsilon_1 \mbox{ of Eq. \ref{FirstOrder2DOFdimless2} for \ref{FirstOrder2DOFmatrixAdef}})
\end{array}%
\end{equation}

For subsequent nonlinear calculations assume $u = 1$ so that contact is active; there is no gap between the HRA and the hull.  Assume $ \epsilon > k $ (dimensionless Equation \ref{FirstOrder2DOFdimless2}) and for stability $ \epsilon_1 k_1 > \epsilon_2 k_2 = k_{\mbox{\scriptsize{base}}} $ (dimensioned 2 DOF Equation \ref{DEfor2DOF}).  Note that all epsilons are dimensionless but they also are a part of the control law that modulates the dimensioned system, Equation \ref{DEfor2DOF}.

Figure \ref{realEigen2dof} on page \pageref{realEigen2dof} plots the real part of values of the eigenvalues.  DOF 1 and 2 represent the hull and armor, respectively, in the 2DOF system.  After a critical $\epsilon$, the higher frequency mode (the clattering, antisymmetric mode) separates from the symmetric mode.  When the hull has a stiffness that sufficiently exceeds the critical stiffness, the symmetric mode settles into a constant eigenvalue with respect to $\epsilon$ regardless of the dimensionless stiffness between the two masses.  ($\partial \lambda_1 / \partial \epsilon \rightarrow 0$ and $\partial \lambda_1 / \partial k \rightarrow 0$ but not for $\lambda_2$.)  From the 3-D FEA results in Figure \ref{impactResponseFRF} on page \pageref{impactResponseFRF}, the higher frequency clattering mode is the lower energy mode.  This is consistent with experience and the Zienkiewicz quote on page \pageref{quote:Zienkiewicz} \cite[340-241]{book:Zienkiewicz91}.  The excitation an antisymmetric mode requires more energy for the same deflection amplitude as an equivalent symmetric mode.  If there is an avenue for vibration strain and strain energy to flow into a lower frequency mode the energy ``conduit" can be a joint or structural connection (nonlinear structural components).  Plucking a taut cord close to the held end, to vibrate it at higher frequency will similarly excite the lowest frequency mode given enough time.

% Commented 25d17 mk
%The results of this subsection are used in the next section in a copy of the Mathematica results \cite{kobold06} for a nonlinear ``eigenvalue" relation for the two DOF problem.  Specifically, the related eigenvectors from Equation \ref{MathematicaSDOFeigenvectors} used in the calculation \cite[360 ff]{Brogan} of a symmetric response $ \Phi_{symm} = (\Phi^T + \Phi)/2 $ provide a set of eigenvalues from $ \Phi^t \Lambda \Phi $ that have the same behavior, thus providing a necessary validation for these results.

%Application of the SDOF model over all DOF's using a theory of linear structural response gives a relationship for a linear transfer function \cite{book:BendatPiersol93} (the Fourier transform of the impulse response \cite{book:ChenSlinSys}).  $ K_i $ is the generalized stiffness for DOF number \textit{i}.  Each of the many DOFs in a linear time-invariant (LTI) system has a relationship that has the same form as shown in a transfer function \cite[p. 81]{MSthesis:kobold06}.  But these functions are often inapplicable.  Common vehicles are filled with nonlinear joints, all of which can be adequately modeled in FEA with some effort. For the ensuing 2 DOF system, the results of the prior subsections and the 2 DOF extension of the SDOF theory are formulated for entry into Mathematica.  This subsection shows the reformulation and the
%Mathematica results.

%% Compressed these paragraphs over this omitted sentence.
%%impulse response \cite{book:ChenSlinSys}).
%%
%%%$ K_i $ is the generalized stiffness for DOF number \textit{i}.

Application of the SDOF model over all DOF's using a theory of linear structural response gives a relationship for a linear transfer function \cite{book:BendatPiersol93} (the Fourier transform of the impulse response \cite{book:ChenSlinSys}).  Each of the many DOFs in a linear time-invariant (LTI) system has a relationship that has the same form as shown in a transfer function \cite[p. 81]{MSthesis:kobold06}.  But these functions are often inapplicable to such nonlinear systems.  Common vehicles are filled with nonlinear joints, all of which can be adequately modeled in FEA, with some effort to avoid misapplication\footnote{Bolted joints do not have the same load paths  as the usual easy way to stitch an FE model, welded axles (rotation only).  This common mistake is usually driven by schedule concerns but it provides answers that are orders of magnitude in error, even for ``loads models."} of St. Venant's principle \cite{book:love1944,book:love2013treatise,ppr:mises1945saint}.  For the ensuing 2 DOF system, the results of the prior subsections and the 2 DOF extension of the SDOF model are formulated with Mathematica.

% Commented 25d17 mk
%The FE model used the simple `non-adaptive' MacNeal-Schwendler NASTRAN CGAP element. This work assumes the closed stiffness is the sum of the two stiffness functions. The contour plot in the thesis \cite[Fig 57]{kobold06} shows that a `base' stiffnesses, $k_{\mbox{base}}$ (vehicle structural stiffness inboard of the armor or skin) is higher than approximately $ 2.4 \epsilon $ that will provide imaginary parts to the eigenvalues.  Positive and negative real parts will cause growth or decay in the state space \cite{ChenSlinSys} (phase space) `orbits' \cite[Fig 12]{kobold06}. Due to nonlinearity, that change is not monotonic.  For example, the response could be oscillatory, growing, or decaying every other half cycle.

\subsection{Eigenvalues: Damped 2 DOF sprung mass}

Repeated roots occur for the eigenvalues $\lambda(\epsilon, k)$ plotted in Figure \ref{realEigen2dof} on page \pageref{realEigen2dof} when the dimensionless closed stiffness is less than a critical stiffness, $ \epsilon < k_{crit} $.  Repeated roots indicate the symmetric and antisymmetric modes have the same frequency.   However, since there are real and imaginary parts to the eigenvalues \cite[171]{book:Thomson88}, that there is a growth in energy for one mode and a decrease in energy for the other.  For most joints that undergo cyclic loads the contact frictional footprint area also oscillates, which causes the stiffness to oscillate.  Therefore, the imaginary parts of the eigen-frequencies act to re-balance the strain-energy according to the control law $u(t)$.  Over time energy will move from the higher energy antisymmetric mode to the lower frequency symmetric mode, unless the lower mode is suppressed.

%\cite[171]{book:Thomson88}, that there is a growth in energy for one mode and a decrease in energy for the other.  During the cycling of loads on most joints, the contact frictional footprint area also oscillates, which varies the stiffness.  Therefore, the imaginary parts of the frequencies act to re-balance the strain-energy.  The control law $u(t)$ approximates this physics.  Over time energy will move from the higher energy antisymmetric mode to the lower frequency symmetric mode, unless it is suppressed.
%
%However, the low base stiffness situation (repeated roots) merely indicates that the stiffness of the base is so low that the close and far masses (a one -- dimensional lumped mass model for the hull and armor) vibrate together as if in free space.  Either they start with negligible relative motion where both together vibrate away from and towards the base (the trivial solution for low base stiffness) where the center of mass is oscillating according to the small base stiffness ($ f_{\mbox{CG}}^2 = k_{\mbox{base}} / 4\pi^2[m_{\mbox{hull}} + m_{\mbox{HRA}}] $),  or they vibrate apart and together with their center of mass remaining stationary.  The eigenvalue increase for anti-symmetric modes when gap stiffness increases provides a basis for similar stiffness-related results seen in the unsymmetrical FEA modes. Additionally, \textbf{this nonlinear eigenvalue behavior indicates that certain modes (`symmetrical') are better target identification features than others} if the deflection at higher frequencies is too small for adequate modulation of the probe beam.

The repeated roots also show that when $k_{base}$ is small enough, the two masses that represent the HRA-hull system vibrate as if they were attached and in free space.  In this case the two masses vibrate together, toward and away from the base, with negligible relative motion.  For this small $k_{base}$ trivial solution the center of mass is oscillates according to  $ f_{\mbox{\scriptsize{CG}}}^2 = k_{\mbox{\scriptsize{base}}} / 4\pi^2[m_{\mbox{\footnotesize{hull}}} + m_{\mbox{\scriptsize{MRA}}}] $.  This is the symmetric mode.  Alternately, they can vibrate apart and together with their center of mass remaining stationary. For such antisymmetric modes the increase in the eigenvalue that occurs when gap stiffness increases provide a basis for similar results seen in the unsymmetrical FEA modes for stiffer bolts and batting material.  The hull and armor 2DOF model will allow DOF 1 and 2 to have different unsymmetrical frequencies sin in general $ m_{\mbox{\footnotesize{hull}}} \neq m_{\mbox{\scriptsize{MRA}}}$.  Where the deflection at higher frequencies does not modulate of the probe beam as well as lower frequencies, \textbf{this nonlinear eigenvalue behavior has the result that symmetrical modes are better target identification features} than the unsymmetrical modes.

\subsection{1-D eigenvectors compared to 3-D FEA}

\label{1dto3dCompared:section7pt6}

%The Mathematica screen print in Figure \ref{MathematicaScreenPlot}
%on page \pageref{MathematicaScreenPlot}

%This DE (Equation \ref{FirstOrder2DOFdimless2}) and the SEVP (Equation \ref{FirstOrder2DOFmatrixEq}) will be shown to be the `special' eigenvalue problem system matrix $ [A_1] $ for a simplified nonlinear contact system.  The system matrix $ A_1 $ provides the special EVP where $ [A_1] \overrightarrow{\Lambda}_i = zero $ for each eigenvector $ \overrightarrow{\Lambda_i} $ in the matrix of (two) eigenvectors $ \mathbf{\Lambda} $.  (This is still a 1-D system, but with 2 DOF, hull deflection and armor deflection.)  There are three terms in the dynamical DE (such as Equation \ref{dimlessDEfor2DOF}) but we can see that the system matrix is [A] from $ [A_1] = |[A]-\lambda [I]| $.

Equations \ref{FirstOrder2DOFdimless2} and \ref{FirstOrder2DOFmatrixEq} introduced the SDOF DE and its `special' eigenvalue problem (SEVP).  In this SEVP the system matrix $ [A_1] $ models a simplified 1-D nonlinear contact system.  The derivation of the system matrix $ [A_1] $, from $[A]$ is on page \pageref{FirstOrder2DOFmatrixAdef} in equations \ref{FirstOrder2DOFdimless2}-\ref{FirstOrder2DOFmatrixAdef}.  The solution to the SEVP $ [A_1] \mathbf{\Lambda} = 0 $ provides two eigenvectors $ \overrightarrow{\Lambda_i} $ from $ \mathbf{\Lambda} $.

This is the solution of the dynamical DE in Equation \ref{dimlessDEfor2DOF}.  It uses \textbf{four main assumptions} \cite{MSthesis:kobold06} to model the contact point for 1-D as a surface of slideline contacts in NASTRAN\texttrademark:  (1) The damping is uniformly constant per unit area and equivalent to the 1-D model, (2) for both the foundation base to the hull as with the hull to the HRA armor.  (3) The combination of low but sufficient base stiffnesses (``open" stiffness) and high ``closed" stiffness is stable.  (4) For the analytical results, a seemingly severe assumption sets dimensionless frequencies to be equal, $\psi_1 = \psi_2 $. The FEA more properly analyzes damping and simulates these results via transient nonlinear time-integrated results using Newmark-Beta methods for algorithmic stability \cite{book:Zienkiewicz91, misc:aero51xAnderson93}, for part of the solutions for the proper 3-D dimensioned DEs for all modes.  Therefore, the millions of 3-D DOF in the FEA remove this last assumption.

In 1-D each of the two DOFs obey the assumed nonlinear solution in Equation \ref{nonlinearSolution} where the amplitude is $ x_{\mbox{\scriptsize{max}}} $ which is scaled by a characteristic length $ L^\ast $ providing a dimensionless solution $x(t)$.  $ L^\ast $ is best chosen by using the Buckingham $\pi$ theorem \cite{misc:aero51xAnderson93, ppr:BuckinghamPi} could range between the average of the microscopic roughness height (microns) to the gap averaged over the surface area in the 3-D model where some part of the HRA-hull system is barely touching.  The latter is most easily be accomplished using the normal modes along with a few pertinent static deformed shapes.

%Also, with this statement, we are saying that the particular solution is the same for both DOF's which is not actually the case, except for purely symmetric and antisymmetric modes.  However, we know that physically symmetric and antisymmetric modes, specifically these two modes alone, comprise the complete set of time solutions in Equation \ref{nonlinearSolution} for the 2DOF undamped SEVP considered. Hence, the assumption $\psi_1 = \psi_2 $ gets validation from the physical dynamics of this two DOF undamped system.  Damping will add a complication to this system but to first order assume both masses have the same particular order of solution, set for each time step.

\begin{equation} \label{nonlinearSolution}
\begin{array}{r}
x(t) = \overline{\frac{x_{\mbox{\scriptsize{max}}}}{L^\ast}} \bigg(\cdots\mbox{nonlinear
terms}\cdots\bigg) e^{j\psi t + \phi_0} %
\qquad \\
\\
\qquad \qquad = a(t)e^{-\mu_i t}\cos(\phi_i t
+ \beta(t))
\end{array}
\end{equation}

%$ \qquad = a(t)e^{-\mu_i t}\cos(\phi_i t
%+ \beta(t)) $

The single DOF solution in equation \ref{nonlinearSolution} applies to both DOF's in matrix form, with cross-terms, in the dimensionless system matrix from Equations \ref{FirstOrder2DOFdimless2}-\ref{FirstOrder2DOFmatrixAdef} on page \pageref{FirstOrder2DOFmatrixAdef} as shown in Equation \ref{systeMatrix2DOF} below.

\begin{equation} \label{systeMatrix2DOF}
[A_1] =  %
\\
\left(%
\begin{array}{cc}
  -\mu^2 - \psi^2 + 1 + \epsilon & -1 \\
  -1 & -\mu^2 - \psi^2 + 1 + k_{\mbox{\scriptsize{open}}} \\
\end{array}%
\right)
\end{equation}

%\textbf{The dimensionless frequencies, $\psi $, just happened to be in this system matrix in the same form as an eigenvalue in the form $ A_1 = |[A] - \psi^2 [I]| $.  This form where $\psi_1^2 =\psi_2^2 = \lambda_i $  are the eigenvalues, only applies to the assumptions of equal damping and frequency for the two masses.} Usually there is a superposition of both symmetric and antisymmetric modes so the hull -- HRA subsystem will resonate at both $ \psi_1 = \pm \sqrt{\lambda_1} $ and $ \psi_2 =
%\pm \sqrt{\lambda_2} $ in a linear combination of modes.  Each of the two $ \Lambda_i $ modes will have both $\psi_1$ and $\psi_2$ active for that one mode. Therefore, the energy applicable to both $\psi_1$ and $\psi_2$ for $ \lambda_1 $ is the same.  This energy will generally be different from that for both $\psi_1$ and $\psi_2$ for $ \lambda_2 $.

\textbf{The dimensionless frequencies, $\psi $, in this system matrix occur in the same form as an eigenvalue $\lambda$ from $ [A_1] = |[A] - \lambda [I]| $.  This form where $\psi_1^2 =\psi_2^2 = \lambda_i $  are the eigenvalues, only applies to the above assumptions of (2) equal damping and (4) equal frequencies for the two masses} which are necessary for a practical nonlinear solution. There is a physical superposition of both symmetric and antisymmetric modes at the same time, resulting in hull -- HRA resonation at both $ \psi_1 = \pm \sqrt{\lambda_1} $ and $ \psi_2 = \pm \sqrt{\lambda_2} $, in a linear combination of modes.  Each of the two $ \Lambda_i $ modes will have both $\psi_1$ and $\psi_2$ active for that one mode. Therefore, $\psi_1$ and $\psi_2$ represent the same energy for $ \lambda_1 $ (symmetric vibration), but $\psi_1$ and $\psi_2$ produce different energy for $ \lambda_2 $ (unsymmetric vibration).

%Therefore, the energy applicable to both $\psi_1$ and $\psi_2$ for $ \lambda_1 $ is the same.  This energy will generally be different from that for both $\psi_1$ and $\psi_2$ for $ \lambda_2 $.

The FEA eigenvalues (i.e., values of the diagonal NASTRAN $ \Lambda $ matrix) are related to their eigenvectors $ \overrightarrow{u_i}^T $, as plotted in the thesis \cite[Fig21-23]{MSthesis:kobold06}. The FEA modal frequencies are the square roots of the eigenvalues, summarized therein \cite[Table 7, Fig 20]{MSthesis:kobold06}. Those frequencies and plots of $ \overrightarrow{u_i}^T $ represent eigenvalues {$ \lambda_i $} within $ [\Lambda] $ and eigenvectors $ \Gamma_i $ within $ [\Sigma] $, in analogy to the 2DOF model, where the columns of $ [\Sigma] $ are the eigenvectors.  As discussed above, the similar-frequency argument makes physical sense, but the similar damping assumption is an artifice used to obtain a practical, simplified solution that is useful for comparisons.  The simplification allows the extra $\psi$ terms to cancel.  Experience and the FEA results both validate that this technique is adequately appropriate for use in this particular case.

%\begin{equation} \label{MathematicaDE}
%[A_1] = \left(%
%\begin{array}{cc}
%  1 + \epsilon -\mu^2 - \psi^2 & -1 \\
%  -1 & k -\mu^2 - \psi^2 \\
%\end{array}%
%\right)
%\end{equation}

%The eigenvalues in matrix form are $[\Lambda] = [ \lambda_1, 0; 0, \lambda_2 ] $ but it is more convenient to display them in the vector form as in Equation \ref{MathematicaSDOFeigenvalues}, which is Equation \ref{FirstOrder2DOFeigenvalues}, changed to reflect the subsequent eigenvector term.

The eigenvalues in matrix form are $[\Lambda] = [ \lambda_1, 0; 0, \lambda_2 ] $ but it is more convenient to display them in the vector form (Equation \ref{MathematicaSDOFeigenvalues}), where the parameter $\kappa$ was  defined in Equation \ref{FirstOrder2DOFeigenvalues}.

\begin{equation} \label{MathematicaSDOFeigenvalues}
\left(%
\begin{array}{c}
  \lambda_1 \\
  \lambda_2 \\
\end{array}%
\right) = \frac{1}{2}\left(%
\begin{array}{c}
  1 + \epsilon + k - \sqrt{\kappa} - 2 \mu^2 - 2 \psi^2 \\
  1 + \epsilon + k + \sqrt{\kappa} - 2 \mu^2 - 2 \psi^2 \\
\end{array}%
\right)
\end{equation}

% Little kappa already defined two equations prior, FirstOrder2DOFeigenvalues}.
%$ \qquad \quad \kappa \equiv 5 - 2 \epsilon + \epsilon^2 + 2 k - 2 \epsilon k - k^2 $

The 2 DOF eigenvectors in Equation \ref{MathematicaSDOFeigenvectors} are one dimensional mode shapes for displacement of the hull, $u$, and the armor, $v$, for modes 1 and 2:

\begin{equation} \label{MathematicaSDOFeigenvectors}
\Sigma = \left(%
\begin{array}{cc}
  u_1 & v_1 \\
  u_2 & v_2 \\
\end{array}%
\right) = \frac{1}{2}\left(%
\begin{array}{cc}
  -1 - \epsilon + k + \sqrt{\kappa} & \quad 1 \\
  -1 - \epsilon + k - \sqrt{\kappa} & \quad 1 \\
\end{array}%
\right)^T
\end{equation}

\subsection{Analysis of the 2 DOF SEVP DE}

\label{analysisOf2DOFassumptions}

%Mathematica\texttrademark shows eigenvalues $ \lambda_i = \omega_i^2 $ for each column $ \Gamma $ of the eigenvector matrix ($ \Sigma $).  Equation \ref{eigenVals1} represents a fixity extreme, a low frequency ``DC" limit $\psi = 0$.
% shows

The eigenvalues $ \lambda_i = \omega_i^2 $ for each column $ \Gamma $ of the eigenvector matrix ($ \Sigma $) are the result of solving Equation \ref{systeMatrix2DOF}.  Equation \ref{eigenVals1} represents an extreme fixity of this Mathematica\texttrademark solution, a low frequency ``DC" limit $\psi = 0$.

% Commented 25d17 mk
%$ \Sigma $ appears on the last two output `lines.'  Equation \ref{eigenVals1} represents a fixity extreme, the low
%frequency ``DC" limit ($\psi = 0$).

%Then the eigenvector matrix (\symbol[ -- Eigenvector matrix]{$
%\Gamma $}) appears on the last two `lines' of regular font output
%for the twoDOFeigen.nb model.

\begin{equation} \label{eigenVals1}
\lim_{\psi=0} \overrightarrow{\lambda} = \frac{1}{2} \left(%
\begin{array}{c}
  1 + \epsilon + k -2\mu^2 - \sqrt{\kappa} \\
  1 + \epsilon + k -2\mu^2 + \sqrt{\kappa} \\
\end{array}%
\right)
\end{equation}

The parameter $\kappa$, defined in \ref{FirstOrder2DOFeigenvalues}, in this first order correction to the linear eigenvalues (defined for Equation \ref{MathematicaSDOFeigenvalues}) is only a function of the stiffnesses, the square of the base and hull stiffnesses.  Compared to prior Mathematica\texttrademark results, this $\psi=0$ model provides a ``DC"  eigenvalue solution that is otherwise not available.

\begin{equation} \label{eigenVectors1}
\Sigma^T =
\frac{1}{2} \left(%
\begin{array}{cc}
  \overrightarrow{\Gamma_1} & \overrightarrow{\Gamma_2} \\
\end{array}%
\right)^T = \frac{1}{2} \left(%
\begin{array}{cc}
  -1 - \epsilon + k + \sqrt{\kappa} & \quad 1 \\
  -1 - \epsilon + k - \sqrt{\kappa} & \quad 1 \\
\end{array}%
\right)
\end{equation}

\textbf{This relation is only valid for the assumptions described earlier: (2) $\mathbf{\mu_1 = \mu_2}$ and (4) $\mathbf{\psi_1 = \psi_2}$.} Otherwise more nonlinear terms remain and the system is not susceptible to SEVP solution for modes in Equation \ref{eigenVectors1}.

\subsection{Synthesis of the 2 DOF SEVP DE}

\label{SEVP2DOFdiscussion}

%% Commented on 25d17 mk
%The synthesis equation for these Mathematica analysis results is Equation \ref{synthesize2DOF2} below.  First we can synthesize the `system' matrix to which these eigenvectors belong.  This subsection is merely a convenient explanation of how these eigenvalues came to be.  The kernel of the SEVP is $ [A_1] = |[A]-\lambda [I]| $.

%First we can synthesize the `system' matrix to which these eigenvectors belong.  The kernel of the SEVP is $ [A_1] = |[A]-\lambda [I]| $. Using the fact that $\psi^2$ is the dimensionless frequency, the approximate system matrix A is available from the terms of $ [A_1] $, defined for Equation \ref{MathematicaSDOFeigenvalues}. This is only approximate because of the many combinations of nonlinear and approximately linear variables (like this `linear' $\psi$) selected in Dr. Major Winthrop's dissertation \cite{phdth:Winthrop04} and used in an available ``Mathematical Preliminaries" section \cite{MSthesis:kobold06}.  Taking the matrix $ [A_1] $ we can add a diagonal of $ \lambda_i [I] = \psi_i^2 [I] $ to extract \textit{A} from $ [A_1] = |[A]-\lambda [I]| $.

This subsection describes the use of Mathematica\texttrademark to synthesize the eigenvalues and eigenvectors from the system matrix [A].  Since the kernel of the SEVP is $ [A_1] = |[A]- \lambda [I]| $, and the eigenvalue solutions are the dimensionless frequencies squared $\lambda_i = \psi_i^2$, the approximate system matrix [A] derives from $ [A_1] $, as defined for Equation \ref{MathematicaSDOFeigenvalues}.  This is only approximate because of the many combinations of nonlinear and approximately linear variables (e.g. ``linear'' $\psi$) selected in \cite{phdth:Winthrop04} and used the ``Mathematical Preliminaries" section of \cite{MSthesis:kobold06}.  Adding a diagonal of $ \lambda_i [I] = \psi_i^2 [I] $ to the matrix $ [A_1] $ allows the extraction of [\textit{A}] from $ [A_1] = |[A]-\lambda [I]| $ as shown in Equation \ref{synthesize2DOF1}.

\begin{equation} \label{synthesize2DOF1}
[A] = \left(%
\begin{array}{cc}
  1 -\mu^2 + \epsilon & -1 \\
  -1 & 1 -\mu^2 + k_{\mbox{\scriptsize{open}}} \\
\end{array}%
\right)
\end{equation}

Equation \ref{synthesize2DOF1} represents a 2 DOF system with springs of stiffnesses k and $\epsilon$, and uniform damping of the same magnitude for both DOFs.  The eigenvectors and eigenvalues are most easily recognized in relation to the standard EVP in Equation \ref{synthesize2DOF2} assuming ($ \lambda_i = \psi^2 \quad \forall \quad i \in [1, 2] $).

\begin{equation} \label{synthesize2DOF2}
\left(%
\begin{array}{cc}
  1 -\mu^2 + \epsilon & -1 \\
  -1 & 1 -\mu^2 + k_{\mbox{\scriptsize{open}}} \\
\end{array}%
\right) \times \overrightarrow{\Gamma_i} %
\\
 = \lambda_1 \times
\overrightarrow{\Gamma_i}
\end{equation}

%% Commented 25d17 mk
%$  \qquad \qquad \qquad \qquad \qquad i \in [1, 2] $

\subsection{Synthesis of unmatched DE}

%% Commented-out 25d17 mk
%Extrapolating the derivation of the eigenvalue solution back to the initial structural D.E. we can use Equation \ref{synthesize2DOF2} and the expressions derived above, the
%solutions in Equations \ref{eigenVals1} and \ref{eigenVectors1} to the SEVP.

%% Commented-out 25d17 mk
%For clarification and to bring us back to the full nonlinear D.E., with the proviso that the eigenvalues for each DOF are the same for each mode as described above; the full nonlinear D.E. with separate damping appears in Equation \ref{synthesize2DOFphysical1}, further below. Except here we use the physical argument

For clarification and to bring us back to the full nonlinear DE, stipulate that the eigenvalues for each DOF are the same for each mode $  \lambda_1 \equiv  \lambda $, that the physical argument leading to Equation \ref{nonlinearSolution} holds the frequencies equal.  Therefore, the frequency subscript matches the subscript for the eigenvector, rather than matching the damping modes as was done in the prior subsection; the combinations of frequency and damping are $(\psi_k, \mu_i) \quad \forall \quad i\ne k$.

The control law, $u(\xi_1)$ of Equation \ref{controLawFor2DOF} is the nonlinear stiffness \cite[Fig 11]{MSthesis:kobold06}, based on the relative displacement, $ \xi_1 = L^\ast x_1 $. The arctangent switch changes the stiffness between the hull lumped point mass and the HRA plate lumped point mass, $ k_{\mbox{\scriptsize{open}}} \leftrightarrow k_{\mbox{\scriptsize{closed}}} $.

\begin{equation} \label{controLawFor2DOF}
u(\xi_1) = \frac{1}{2}- \arctan(k_x \xi_1)
\end{equation}

The phase $ \phi_k = \psi_k t  +  \beta(t) \approx \psi_k t $  relates to the eigenvector whose temporal dynamics $ \lambda_i $ describes both DOFs, the hull and HRA point masses. Here the frequencies $ \psi_k $ are not matched to the damping $ \mu_i $.  Redefining $\Pi_i$ and $\Omega_i$ for this mixed-index format:

%$ \forall \quad \Pi_{i,k} \equiv (\mu_i^2 - \psi_k^2) + 2\mu_i\psi_k\arctan \phi_k,$
%
%$ \qquad \Omega_{i,k} \equiv 2\mu_i \bigg[-\mu_i - \psi_k\arctan \phi_k \bigg] $

\begin{equation} \label{synthesize2DOFphysical1}
\begin{array}{l}
\forall \quad \Pi_{i,k} \equiv (\mu_i^2 - \psi_k^2) + 2\mu_i\psi_k\arctan \phi_k \\
\qquad \Omega_{i,k} \equiv 2\mu_i \bigg[-\mu_i - \psi_k\arctan \phi_k \bigg] \\
\\
 \qquad \quad \left(%
\begin{array}{cc}
  \Pi_{1,k} & 0 \\
  0 & \Pi_{2,k} \\
\end{array}%
\right) \times \left(%
\begin{array}{c}
  \Lambda_1 \\
  \Lambda_2 \\
\end{array}%
\right)_k + \\
 \qquad \quad \left(%
\begin{array}{cc}
  \Omega_{1,k} & 0 \\
  0 & \Omega_{2,k} \\
\end{array}%
\right) \times \left(%
\begin{array}{c}
  \Lambda_1 \\
  \Lambda_2 \\
\end{array}%
\right)_k  + \\
  \left(%
\begin{array}{cc}
  1 + \epsilon u(x_1) & -1 \\
  -1 & k_{\mbox{\scriptsize{open}}} \\
\end{array}%
\right) \times \left(%
\begin{array}{c}
  \Lambda_1 \\
  \Lambda_2 \\
\end{array}%
\right)_k  = \left(%
\begin{array}{c}
  0 \\
  0 \\
\end{array}%
\right) \\
\end{array}
\end{equation}

Equation \ref{synthesize2DOFphysical1} provides an independent identically distributed (iid) estimate of the dynamics modeled as $N$ iid oscillators.  Since oscillators for each DOF in a FE model and in continuous media (reality) are not independent nor of a random distribution in a structure, this relationship merely guides independent behavior that quickly effects other DOF's.  More importantly, the ``nonlinear" eigenvalue behavior in Figure \ref{realEigen2dof} on page \pageref{realEigen2dof} shows how the symmetrical modes are affected by stiffness change \cite[p. 148]{MSthesis:kobold06}.

%A Lyapunov function form of energy balance defined for this nonlinear contact system \cite[App F]{MSthesis:kobold06} serves to help validate the stability of SDOF and 2DOF relations.

A Lyapunov function of energy balance defined for this nonlinear contact system in the thesis \cite[App F]{MSthesis:kobold06} validates the stability of these composite SDOF and 2DOF relations.

\bibliographystyle{ieeetran}

% \bibliography{plain, modLobsrv}   % Seems like IEEEtran.bst defines two arguments, plain one.

% \bibliography{modLobsrv}

\bibliography{modLobsrvFinal}

%\bibitem{}
%, \textit{}

{}
%
%\end{thebibliography}

\end{document}